\documentclass [10pt,preprint2]{aastex}
\begin{document}

\slugcomment{2-nd submission to The Astrophysical Journal}
\slugcomment{August 27, 2001}
\title{Emission-Line Properties of $z > 4$ Quasars \altaffilmark{1,2}} 

\altaffiltext{1}{Based in part on observations obtained at the MMT
Observatory, a joint facility of the University of Arizona and the
Smithsonian Institution.} 
\altaffiltext{2}{Based in part on observations obtained at the
W.M. Keck Observatory, which is operated as a scientific partnership
among the California Institute of Technology, the University of
California and the National Aeronautics and Space Administration.  The
Observatory was made possible by the generous financial support of the
W.M. Keck Foundation.}

\author{Anca Constantin and Joseph C. Shields}
\affil{Department of Physics and Astronomy, Ohio University, Athens, OH 45701}
\author{Fred Hamann}
\affil{Department of Astronomy, University of Florida, Gainesville, FL 32611}
\author{Craig B. Foltz}
\affil{MMT Observatory, Tucson, AZ 85721}
\author{Frederic H. Chaffee}
\affil{W. M. Keck Observatory, Kamuela, HI 96743}

\begin{abstract}

High redshift QSOs provide important opportunities for advancing our
understanding of the astrophysics of galaxy formation and evolution.
Although a growing number of these sources are now known to exist at
redshifts beyond 4, systematic studies of the emission-line properties
of these objects are quite limited.  We present results of a program
of high signal-to-noise spectroscopy for 44 QSOs using the MMT and
Keck observatories.  The quasar spectra cover 1100 -- 1700 \AA{} in
the rest frame for sources spanning a luminosity range of $\sim 2$
orders of magnitude. Comparisons between these data and spectra of
lower redshift quasars reveal a high degree of similarity, although
differences are present in the profiles and the strengths of some
emission features. An examination of the luminosity dependence of the
emission lines reveals evidence for a weak or absent Baldwin effect
among $z \ga 4$ QSOs.  We compare measurements for objects in our
sample with results from other high redshift surveys characterized by
different selection techniques. Distributions of equivalent widths for
these different ensembles are consistent with a common parent
population, suggesting that our sample is not strongly biased, or in
any case, subject to selection effects that are not significantly
different from other surveys, including the Sloan Digital Sky Survey.
Based on this comparison, we tentatively conclude that the trends
identified here are representative of high $z$ QSOs. In particular,
the data bolster indications of supersolar metallicities in these
luminous, high-$z$ sources, which support scenarios that assume
substantial star formation at epochs preceding or concurrent with the
QSO phenomena.

\end{abstract}


\section{Introduction}

The properties of QSOs at the highest known redshifts are important
for understanding the early history of galaxies and the formation of
supermassive black holes.  Over the past decade, a growing number of
quasars have been reported at redshifts $z > 4$, corresponding in
look-back time to more than $\sim 90$\% of the age of the Universe.
Objects in this regime are thus inherently young, and accordingly of
great interest for studying the onset of accretion-powered activity in
relation to galaxy formation and evolution.

Optical studies of QSOs at $z \gtrsim 4$ to date have mostly
emphasized their discovery, luminosity function, and use as probes of
intervening matter seen in absorption.  Investigations
of the emission-line properties of these objects have been far more
limited \citep{sch91,sto96}.  Further scrutiny of their
emission spectra in relation to those of their low-redshift counterparts
is clearly desirable.  Specifically, such comparisons may reveal
information on the timescale of star formation and chemical evolution
associated with the QSO phenomenon \citep{ham99}, and on time-dependent
behavior of the accretion process \citep{mat00}.

This paper presents a detailed analysis of the optical emission line
properties for a large sample of $z \ga 4$ QSOs. High signal-to-noise,
moderate resolution spectra of 44 quasars were obtained over multiple
observing runs at the MMT and Keck observatories. The study employs
direct measurements of the emisson lines as well as comparisons of
composite spectra.  Preliminary results from this survey were reported
by \citet{shi97}.  Here we compare our spectra with those of redshift
$\sim 2$ quasars and find general agreement, although subtle
differences are present in the strengths and profiles of some emission
features. Our objects show little evidence for a Baldwin Effect, the
empirical anticorrelation between line equivalent width and luminosity
$L$ often found in lower redshift samples (Baldwin 1977; see Osmer \&
Shields 1999 for a review).  We also investigate the degree to which
our findings may be influenced by selection effects.  We compare the
properties of our targets with published measurements for other $z \ga
4$ QSO samples representing a variety of discovery methods.  This
analysis indicates that our sources are representative of known
quasars in this redshift regime, and suggests that the observed
differences from lower-$z$ sources are at least partially intrinsic.
In subsequent papers (e.g., Dietrich et al. 2001), we will present
more detailed comparisons of quasar emission-line properties as a
function of redshift and luminosity, based on a sample of $>800$
objects that span a much wider range in $z$ and $L$.

Although the values of the adopted cosmological parameters play an
important role at these extreme redshifts, we adopt throughout the
paper an $H_0 = 50$ km s$^{-1}$ Mpc$^{-1}$, $q_0 = 0.5$, $\Lambda = 0$
cosmology, to ease the comparison with earlier published work in this
field.

\section{Observations \& data analysis} \label{data}

We began a program of spectroscopy of $z \ga 4$ QSOs in 1994, with
targets selected from the sources known at that time ($\sim 50$) or
reported in subsequent years in the literature.  Data were obtained in
a series of 7 observing runs at the MMT Observatory during 1994 --
1996, and 2 runs at the W. M. Keck Observatory in 1997 October and
1999 June.  The goal of this program was to obtain a statistically
significant description of quasar spectra in this high redshift
regime; 32 objects were observed at the MMT, and 12 at Keck, for a
total of 44 QSOs.  Specific targets were selected based on their
observability, with the brightest available sources given priority in
order to maximize the resulting signal-to-noise ratio.  We generally
avoided sources known to exhibit strong absorption lines near the
emission redshift.  The majority of the observed objects are taken
from the APM Color Survey \citep[BR/BRI prefix]{sto96} and the Second
Palomar Sky Survey \citep[PSS prefix]{ken95b}.  Table~\ref{tbl-1}
lists the redshifts, $r$ magnitudes, and the discovery reference for
each source.

Data were acquired with the Red Channel of the MMT Spectrograph
\citep{schm89} and with LRIS at the Keck {\sc ii} telescope
\citep{oke95}.  Observations were obtained with a long slit of
1\arcsec width, yielding full width at half maximum (FWHM) spectral
resolution of $\sim10$ \AA\ and $\sim7$ \AA, with pixel sampling of
3.53 \AA\ pixel$^{-1}$ and 1.86 \AA\ pixel$^{-1}$, respectively. These
resolutions correspond to 300 -- 400 km s$^{-1}$ across the \ion{C}{4}
$\lambda$1549 emission line, or typically 0.07 -- 0.1 times this
line's FWHM.  The spectral coverage was chosen to span the redshifted
Ly$\alpha$ $\lambda$1216 -- \ion{He}{2} $\lambda$1640 interval; the
\ion{He}{2} feature is of specific interest since the \ion{N}{5}
$\lambda$1240/\ion{He}{2} $\lambda$1640 ratio can be used as an
abundance diagnostic \citep{ham93,fer96}.  The observational setup was
essentially identical for the two telescopes.  A long-pass filter
(LP495 at the MMT, GG495 at Keck) was used to prevent second-order
contamination.

Multiple exposures totaling $\sim 1 - 2$ hrs were obtained for each
quasar, yielding a typical signal-to-noise ratio of $\sim 50$ per
pixel.  Observations were obtained at an airmass $\la 1.5$, and the
slit was oriented approximately perpendicular to the horizon in order
to minimize the effects of atmospheric differential refraction
\citep{fil82}.  The atmospheric seeing was typically 1{\arcsec} (or
better for the Keck {\sc ii} telescope), but not all the observations
were obtained under absolute photometric conditions.

\begin{figure}
\figurenum{1}
\plotone{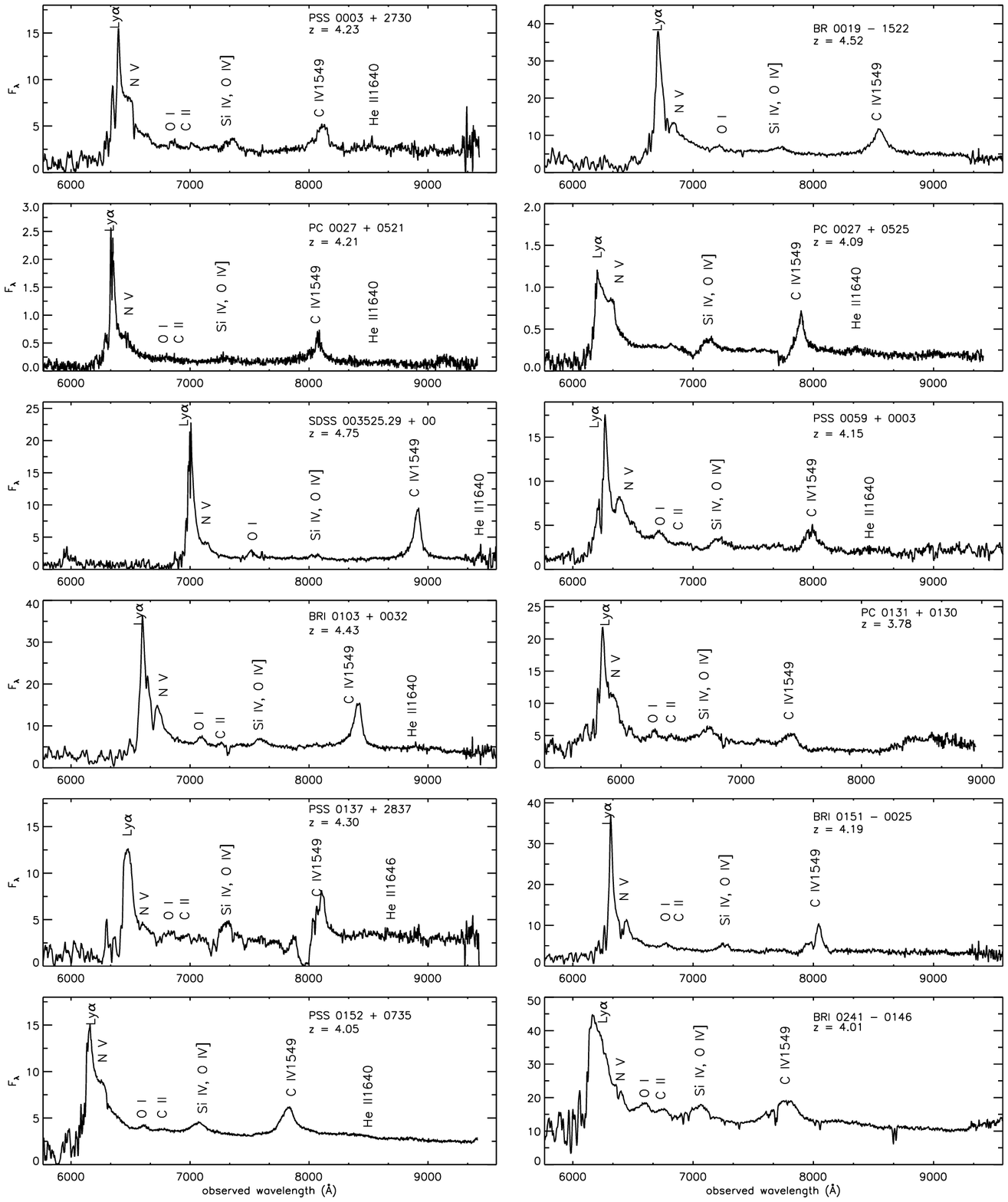}
\caption{Observed spectra of QSOs at $z \ga 4$. The flux, 
F$_{\lambda}$, is in units of $10^{-17}$ erg s$^{-1}$ cm$^{-2}$
\AA$^{-1}$. The prominent emission features are marked.\label{Fig1}}
\end{figure}

Calibration and reduction of the resulting spectra were carried out
using standard techniques as implemented in the IRAF software
package\footnote{The Image Reduction and Analysis Facility (IRAF) is
distributed by the National Optical Astronomy Observatories, which is
operated by the Association of Universities for Research in Astronomy
Inc.  (AURA), under cooperative agreement with the National Science
Foundation.}.  Observations of nearly featureless standard stars were
used to provide flux calibrations and to correct for the atmospheric
absorption bands. He-Ne-Ar arc-lamps were used for the wavelength
calibrations.  Multiple spectra acquired for the same object were
averaged.  Figure~\ref{Fig1} presents all the reduced spectra included
in this observational program, sorted on RA.

A small number of known Broad Absorption Line (BAL) QSOs were observed
as part of the sample in order to enable additional comparisons with
low-redshift quasars. These sources are PSS 1048+4407, BR 1144-0723,
and PSS 1438+2538. Our spectra, which have greater S/N and
wavelength coverage than many of the published discovery spectra, also
reveal for the first time broad absorption lines in PC 0027+0525 and
PSS 0137+2837.

The primary focus of the present work is on the emission lines.
Accurate measurement of these features requires correction for
absorption in many cases.  For the majority of objects, the absorption
appears as narrow lines due to associated or intervening systems,
superimposed on either the emission lines or continuum.  In these
cases we removed the absorption feature through a simple
interpolation.  For the BAL QSOs and several additional objects (BRI
0151-0025, PSS 1159+1337, BRI 1346-0322, PC 1415+3408, and PSS
1435+3057), the absorption was severe enough that a reliable
reconstruction of the unabsorbed spectrum was not possible.  We
excluded these sources from our measurements of the emission lines and
from our composite spectra. All objects in this sample show strong
Ly$\alpha$ forest absorption, which clearly influences the measured
profile and flux of the Ly$\alpha$ emission line in many cases; we did
not attempt to correct for this effect. Objects in our sample are
subject to small amounts of Galactic foreground extinction, spanning
$A_V = 0.04 - 0.49$, with a typical value of $\sim 0.1$ \citep{sch98}.
We corrected the spectra for the resulting reddening using the
empirical selective extinction function of \citet{car89}.

The spectra were Doppler-corrected to the rest frame using
redshift values that we measured for the majority of sources from the
\ion{C}{4} $\lambda$1549 line.  The observed wavelength for each case
was established by fitting a Gaussian to the top 20\%\ of the profile.
For situations where the measured line was significantly affected by
absorption or low signal-to-noise, we verified the result with fits to
the top 50\%\ of the profile.  For the BAL QSOs, the redshifts listed
in Table~\ref{tbl-1} are taken from the discovery reference.  For the
remaining objects where \ion{C}{4} was not measurable due to
absorption (see above), Ly$\alpha$ was used instead.  The
redshift measurements in these cases relied on Gaussian fits that used
as a template only the red side of the profile. The resulting $z$
determinations are in good agreement with previously published values,
with an estimated uncertainty of $\leq 0.01$.

Luminosities for our sample targets were derived using published
photometric data.  Rest-frame 1450\AA\ fluxes (AB magnitudes at
1450\AA) were obtained directly from the discovery papers for the
majority of the objects in our sample. For the objects where these
numbers were not available, the flux measured from our spectra was
used, after the absolute flux calibration was adjusted to obtain
consistency with published broad-band photometry. The luminosity
values, expressed in ergs s$^{-1}$ Hz$^{-1}$, are also included in
Table~\ref{tbl-1}.

To investigate the emission-line properties of the observed QSOs, we
constructed composite spectra, and employed measurements of the lines
in individual objects.

The equivalent widths (EWs) of the strong lines, specifically for
Ly$\alpha$ + \ion{N}{5} and \ion{C}{4} $\lambda 1549$, were measured
interactively, using a direct integration of the line flux referenced
to the interpolated continuum level. For the Ly$\alpha$+\ion{N}{5}
feature we used the extension of the continuum from the red side only,
and did not attempt any deblending.  The resulting values for
\ion{C}{4} are very similar to those obtained by modelling the lines
with 2 Gaussians, representing broad and narrow components. Possible
inaccuracies in these measurements are dominated by ambiguities in
placement of the continuum; experiments with alternative choices of
the continuum level suggest that this contributes a systematic
uncertainty of $\sim 20\%$.

In our analysis, we found it useful to combine our measurements with
published emission-line data for other $z \ga 4$ QSO samples, and in
particular those discovered via the Palomar CCD Grism Survey
\citep[hereafter referred to as SSG, $\sim$ 15
objects]{sch91,sch97,sch95}, and early discoveries from the Sloan
Digital Sky Survey \citep[$\sim$ 35 objects, only those with $z > 3.9$
were considered in this analysis]{fan99,fan00,fan00b}.  In contrast
with our measurements, the published EW values for these samples rely
on single component Gaussian fitting.  Our results and the published
EWs can be meaningfully combined only if the line measurement
procedures are consistent. We tested different methods of measuring
the emission lines with our data. For the \ion{C}{4} line, the single
Gaussian fits give, in general, lower EWs than those obtained with our
direct integration technique, as they tend to lose flux from the wings
if they are prominent, and from the peak if the cores dominate.  In
order to obtain a reasonable consistency in the comparisons involving
the SSG and the SDSS samples, we measured the \ion{C}{4} EWs in both
ways. For Ly$\alpha$ + \ion{N}{5}, the fit results are sensitive to
the method of treating the two lines and the asymmetry resulting from
Ly$\alpha$ forest absorption; consequently, we employed only the EWs
obtained by direct integration of the profile. The values are recorded
in Table~\ref{tbl-1}.

Finally, we constructed composite spectra using the data for individual
sources normalized to unit mean flux over the wavelength range 1430
\AA -- 1470 \AA.  Average and median composites were calculated over
the wavelength range common for all objects.  The average composite
may be influenced by several extreme objects with unusually strong and
narrow lines (BR0401-1711, BRI1050-0000, BR2212-1626,
BR2248-1242). The median spectrum may thus be more representative of the
typical object in our sample.

\section{Spectroscopic properties}
\subsection{General characteristics} \label{general}

An initial scrutiny of the spectra of $z > 4$ QSOs reveals substantial
agreement with those of their lower redshift counterparts.  This
similarity is noteworthy, given the widely differing amounts of time
available for the evolution of the AGN and underlying host galaxy.  In
Figure~\ref{Fig2} (upper panel) we compare our $z \ga 4$ average
composite spectrum with the composite spectrum derived as an
average\footnote{The algorithm used in constructing the LBQS composite
is described by \citet{fra91}} from the total {\it Large Bright Quasar
Survey} (LBQS) sample, which is weighted toward $z \approx 2$ objects
in this wavelength range (S. Morris 1999, private comunication; see
Brotherton et al. 2001 for additional information).  This comparison
is especially useful given that our sample and the subset of the
LBQS contributing to this wavelength interval have nearly identical
average luminosities ($L_{\nu}({\rm 1450\AA}) \approx 31.3$), so that any
differences should reflect a redshift dependence or differing
selection criteria.  We estimated the average LBQS
$L_{\nu}({\rm 1450\AA})$ from the absolute B magnitude
given by \citet{fra91}, their Figure 8, following the
methodology of \citet{sch89} to relate the flux level in the two
different bandpasses, while correcting for differences in
cosmological parameters.  In general terms, the high and intermediate
redshift composites agree remarkably well; they have the same strong
emission lines, similar continuum shape redward of the Ly$\alpha$
feature, and comparable strengths and profiles of the lines. The
pronounced Ly$\alpha$ forest in the $z \ga 4$ composites does not
reflect an intrinsic difference in the nature of the emission sources,
but rather an increase in the opacity of the intergalactic medium at
larger redshifts.

Similarities between the QSOs at $z \ga 4$ and those at lower redshift
can also be seen in measurements of spectral features for individual
sources.  Surveys of QSOs at $z\approx 2 - 3$ \citep{fra92,fra93}
display strong (negative) correlations of line widths (FWHM) with line
EWs and with line peak/continuum ratios. Similar tendencies are
present in the high $z$ sources.  Figure~\ref{Fig3} illustrates these
results for all quasars in our sample, for which the \ion{C}{4} lines
were measurable.  Sources with narrow lines also tend to show the
strongest emission.  Furthermore, spectra of QSOs at $z \la 3$ exhibit
systematic line velocity shifts that correlate with the degree of
ionization of the emitting species \citep{gas82, esp89, tyt92, mci99}. As
reported previously by \citet{sto96}, this behavior is also present in
$z > 4$ samples.

A detailed comparison of the high and intermediate redshift composites
in Figure~\ref{Fig2} shows excellent agreement for some lines but
dissimilarities in others.  \ion{C}{2} $\lambda 1335$ and
\ion{Si}{4}+\ion{O}{4}] $\lambda 1400$ remain basically unchanged,
while differences are present in the strengths of Ly$\alpha$,
\ion{N}{5}, \ion{C}{4}, and \ion{O}{1} $\lambda 1304$.  The Ly$\alpha$
emission in $z \ga 4$ quasars is considerably affected by the forest
absorption, which may remove up to $\sim 50\%$ of the line, but the
high-$z$ composites nonetheless show a stronger Ly$\alpha$ feature
than in the $z \approx 2$ average. The \ion{N}{5} emission is also
enhanced in the high $z$ systems, and the same is true for the
\ion{O}{1} $\lambda 1304$ emission.  The rest-frame equivalent width
of the latter feature is $\sim 3$\AA\, in the composite $z \ga 4$
spectra, which is about 50\%\ larger than that seen in spectroscopic
surveys at lower redshifts \citep[EW(\ion{O}{1}) $\thickapprox
2$\AA]{bro94,for00}.

The \ion{C}{4} line is stronger in the $z \gtrsim 4$ average than in
the intermediate $z$ composite spectrum.  The difference is most
pronounced in the line core, while the line wings are more nearly
comparable in the two composites.  As noted previously, the sample
includes several remarkable objects with unusually strong and narrow
lines, which potentially distort the comparison when using an average
composite for this modest sample. To address this question, the high
$z$ median and the redshift $\sim 2$ composite are shown for
comparison in Figure~\ref{Fig2}, lower panel.  The Ly$\alpha$ and
\ion{C}{4} features are clearly reduced in the high $z$ median, but
still stronger, especially in the core, than those in the LBQS
composite.

Aside from the differences in Ly$\alpha$ and \ion{C}{4}, the high
redshift median and average spectra are remarkably similar.  In
particular, the \ion{O}{1} and \ion{N}{5} emissions are equally strong
in both high $z$ composites.  The degree of variation between the
members of our sample can be seen in the standard deviation of the
normalized spectra as a function of wavelength, plotted in
Figure~\ref{Fig4}.  The largest variations are present in the cores of
the prominent emission lines (Ly$\alpha$ and \ion{C}{4}), consistent
with the previous results for lower redshift samples
\citep{fra92,bro01}.  The lower panel of Figure~\ref{Fig4} shows the
standard deviation of the mean as a function of wavelength, which
provides a measure of the uncertainty of the average spectrum.

\subsection{\ion{O}{1} and \ion{N}{5}} \label{O1N5}

The \ion{O}{1} $\lambda 1304$ and \ion{N}{5} $\lambda 1240$ lines show
differences between the intermediate and high redshift spectrum
composites that merit further attention. These features potentially
convey information about the structure of the broad-line region and
the physical conditions in the emitting plasma. Additionally, both
lines are possible tracers of metallicity in the QSOs, which is of
particular interest for understanding these sources in an evolutionary
framework.

We first examine whether we have correctly identified the \ion{O}{1}
feature.  Previous quasar emission line studies have raised the
question of whether the $\lambda 1304$ feature is indeed \ion{O}{1}
emission or something else.  The \ion{O}{1} line is a triplet
($\lambda 1302.17, \lambda 1304.86, \lambda 1306.03$) with a mean
laboratory wavelength (obtained by weighting each transition by its
Einstein A coefficient) of $1303.50$\AA\ \citep{ver96}.  For the high
optical depths that may occur in broad line clouds, the individual
features in the triplet will approach equal strengths, modifying the
mean to $1304.36$\AA.  The feature we attribute to \ion{O}{1} is seen
at a wavelength of $\sim 1307$\AA\ in these composites.  At first
glance, a better identification might seem to be with \ion{Si}{2}
$\lambda\lambda 1304.37, 1309.27$ (mean laboratory wavelength
1307.63\AA).  However, this doublet is expected to be accompanied by
considerable emission in other \ion{Si}{2} lines, such as
$\lambda\lambda$1196, 1264, 1531, 1817 \citep{gas82, dum81}, which are
weak or absent.  Furthermore, we note that the high $z$ composites are
derived from spectra that are Doppler-corrected using redshifts given
by \ion{C}{4}, which is blueshifted relative to the low ionization
lines.  An identification of the $\lambda$1307 feature with \ion{O}{1}
would be consistent with this trend, while \ion{Si}{2} would not.  We
conclude that the \ion{O}{1} $\lambda$1304 identification is secure.

The enhanced strength of \ion{O}{1} in the high redshift sources may
reflect a structural difference in their broad-line regions (BLRs).
Several studies of AGN ensembles suggest that low ionization lines
(e.g. \ion{O}{1}, \ion{C}{2}) are emitted preferentially in
low-velocity BLR components, while Ly$\alpha$ and high ionization
lines (\ion{N}{5}, \ion{Si}{4}, \ion{C}{4}) tend to trace higher
velocity plasma \citep{bro94, fra92, wil93, bal96}.  This is sometimes
described in terms of a BLR with two components: the Intermediate Line
Region (ILR) and Very Broad Line Region (VBLR).  In our high redshift
sample, the narrow component makes a larger fractional contribution to
the spectra than in the lower redshift systems.  The \ion{C}{4} line has
a larger velocity width in the LBQS composite (FWHM $\thickapprox$
5600 km s$^{-1}$ ) than in the $z \ga 4$ composites, both average and
median (FWHM $\thickapprox$ 4170 km s$^{-1}$ and 4900 km s$^{-1}$,
respectively).  The unusual strength of \ion{O}{1} emission may thus
be the most obvious tracer of a characteristically enhanced ILR in $z
> 4$ QSOs.

The strong \ion{N}{5} emission is not expected in systems with an
enhanced narrow component, however, because of its high
ionization. The systematically higher \ion{N}{5} emission may instead
reflect high metallicity, consistent with standard chemical evolution
scenarios for young, massive elliptical galaxies or galaxy
bulges. \citet{ham92,ham93} suggested that \ion{N}{5}, which is a
collisionally excited line, may be elevated relative to other lines as
a consequence of the secondary nitrogen enrichment in vigorously
star-forming environments.  Interestingly, enhancements in the
\ion{O}{1} emission may provide supporting evidence for this
interpretation.  \ion{O}{1} $\lambda 1304$ is apparently formed via
fluorescence with the H Ly$\beta$ line (see, e.g. Netzer 1990 for a
discussion).  The strength of the \ion{O}{1} feature is thus expected
to scale with the O/H abundance ratio \citep{net76}.  However, the
\ion{O}{1} line strength may be influenced by various other factors,
such as the sensitivity of the line coincidence to the velocity field
in the emitting region.

To summarize, the \ion{N}{5} and \ion{O}{1} features suggest that the
broad-line regions of QSOs at $z > 4$ may be intrinsically different
from those at lower redshift.  Low-velocity components of the BLR
are more prominent in these high-$z$ sources, as signaled by the
strength of \ion{O}{1} and the narrow velocity width for \ion{C}{4}.
In addition to this structural difference, however, the strength of
\ion{N}{5} emission implicates characteristically higher metallicities
in the $z \ga 4$ objects.  High metallicities may also contribute to
the enhancement of the \ion{O}{1} line.

\subsection{The EW vs. Luminosity correlation}

To investigate the amplitude of the Baldwin Effect for QSOs at $z \ga
4$, we present plots of the rest equivalent width of \ion{C}{4}
$\lambda 1549$ as a function of the continuum luminosity $L_{\nu}$
(measured at 1450 \AA\ in the rest frame), for our observed sample
(Figure~\ref{Fig5}, upper panel), and for an enlarged sample that
includes also measurements obtained from the SSG and SDSS surveys
(Figure~\ref{Fig5}, lower panel).  The dotted line is the relation
found by Osmer et al. (1994) for a sample of 186 quasars covering a
wide range of redshift ($z < 3.8$).  The overall results at high
redshift ($z > 4$) appear compatible with the trend and scatter found
for AGNs at lower redshift and luminosity, but do not, by themselves,
show a Baldwin Effect correlation in \ion{C}{4}. A Spearman test is
consistent with the null hypothesis at $ > 99\%$ level. The
corresponding diagrams for Ly$\alpha$ + \ion{N}{5} are shown in
Figure~\ref{Fig6}, and likewise are consistent with no correlation.
These measurements combine Ly$\alpha$ and \ion{N}{5}, and therefore, a
possible trend in the Ly$\alpha$ line alone, for which the Baldwin
Effect has been previously claimed, may be weakened by the presence of
\ion{N}{5}, which does not show such $EW - L_{\nu}$ anticorrelation
\citep{osm94,kor98}.

To explore further the luminosity-dependent behavior, we divided the
quasars in our sample into two luminosity bins, with log
$L_{\nu}$(1450 \AA) $= 30.34 - 31.12$ and $31.31 - 31.97$.  Average
and median composites for these subsets are shown in
Figure~\ref{Fig7}.  The noise and the intrinsic peculiarities present
in single sources are considerably diminished in the composite
spectra, making them a valuable tool in identifying line strength
variations originating in differences in luminosity. A simple
inspection of the individual lines reveals a small decrease in the EW
with luminosity for Ly$\alpha$, but not for \ion{C}{4} or other
emission lines.

The weak or absent Baldwin Effect in our $z \ga 4$ sample may be an
intrinsic characteristic of these objects, which are extreme in both
redshift and luminosity. Alternatively, the absence of a clear trend
may stem from the limited luminosity range of our sample and the
substantial scatter that always appears in the $L_{\nu} - EW$
relation.  Selection effects influencing our sample are also a
significant concern; if strong-lined objects are over-represented
among our high luminosity QSOs, the result would be a weakened Baldwin
correlation. We explore the role of selection effects further in the
next section.

\subsection{Selection biases}

An important consideration in interpreting the properties of our
high redshift QSOs is the extent to which selection effects can influence
the properties of a sample.  Here we examine the methods used for
identifying $z \ga 4$ QSOs and ways that selection criteria may
be reflected in emission-line characteristics.

The majority of QSOs known at $z > 4$ have been discovered by
color-selection techniques \citep{irw91, ken95a, ken95b, sto96, sto01} that
rely in particular on the contrast between the observed spectral
region centered on the Ly$\alpha$ feature, and the continuum at
shorter wavelengths, which is strongly absorbed by the Ly$\alpha$
forest.  Quasars are thus distinguished from stars primarily on the
basis of their extremely red B -- R or similar colors, which separate
them from the stellar locus in the color-magnitude/color-color
diagrams.  The emission lines can contribute significantly to the
total flux observed in typical photometric bandpasses. Their role is
amplified at these particularly high redshifts since the scaling
factor ($1 + z$) boosts the values of the observed EWs.

The prominence of Ly$\alpha$ could therefore introduce biases
affecting the detectability of $z \ga 4$ quasars.  Most of the objects
in our sample come from the APM Color Survey (BR/BRI quasars). These
quasars are drawn from a magnitude-limited sample, and as a
consequence, the strong-lined sources could be over-represented
because of the contribution of the Ly$\alpha$ line to the measured
flux in the R band (Malmquist bias).  The amplitude of this effect can
be roughly evaluated by estimating the change in the apparent (R)
magnitude produced by the presence of a strong Ly$\alpha$ line: for a
rest frame EW$_1 = 30$\AA\ (weak line) and EW$_2 = 160$\AA\ (strong
line) with respect to the unabsorbed continuum, the observed ($z
\approx 4.5$) equivalent widths would be EW$^{obs}_1 \approx 165$
{\AA} and EW$^{obs}_2 \approx 880$ {\AA} respectively.  With an R
bandpass of width $\sim 1000$ \AA, the line increases the measured
flux by $\sim$ 16\% and 88\% compared with a pure continuum source.
The ratio of the two broadband fluxes with strong compared to weak
Ly$\alpha$ would be $f_2 / f_1 \simeq 1.62$, which corresponds to a
difference in magnitude $\simeq 0.5$ mag.  This value is actually an
underestimate of the effect because the continuum blueward of
Ly$\alpha$ is highly suppressed by the absorption forest.  In
addition, for some of the observations in the BRI survey, a narrower R
bandpass ($\sim 600$\AA) was used, resulting in an increase of the
fractional contribution of Ly$\alpha$ to the measured flux.  This
effect leads to a bias in favor of objects with large EW, that could
potentially account for the unusual examples of strong-lined QSOs in
our sample.  The same trend was suggested previously by
\citet{ken95b}, who calculated the selection probabilities of QSOs
(from POSS II observations) for three cases of Ly$\alpha$ + \ion{N}{5}
line strength (see their Fig. 6).  These considerations apply to the
BR/BRI, PSS, and SDSS sources, which rely on similar bandpasses for
QSO identification.

Although this effect seems to be important, it could be counteracted
by the treatment of the other bandpasses in the selection process.  If
only sources that are detected in B are included, as is the case for
the APM survey, objects with very strong lines near the limiting R
magnitude may be excluded, because their R-band flux is dominated by
Ly$\alpha$ and the continuum (which dominates in B) is correspondingly
weak.  However, the selection process is further affected by the fact
that, for very weak lines, the contrast between the Ly$\alpha$
emission and the continuum blueward of it may not be high enough to
satisfy the necessary color criterion.  In the case of the APM survey,
the understanding of this bias is complicated by the use of a
magnitude-dependent color threshold, due to signal-to-noise
considerations \citep[see their Figure 1]{sto01}, again introducing a
preference for strong-lined sources.  The fact that the majority of
the color selected objects in our sample fall above the dotted line in
the Baldwin diagram, may be due, in part, to this bias.

A thorough understanding of the Malmquist and color biases would
require extensive modeling based on the choices of filters, limiting
magnitudes, and/or colors.  An alternative is to investigate the EW
properties of samples discovered by different techniques. We compared
the EWs and composite profiles of our color-selected sources to a
grism-selected sample, which includes objects observed by us and by
\citet{sch91}. The SSG sample consists of quasars selected based on
the detection of Ly$\alpha$ in slitless spectra. As discussed by
\citet{sch95}, this detection technique may also lead to preferential
selection of QSOs with strong lines. Figure~\ref{Fig8} shows the
rest-frame and the observed EW distributions, for \ion{C}{4} and
L$\alpha$, for the color-selected and grism-selected samples. The K-S
(Kolmogorov-Smirnov) statistics are consistent with the two samples
having the same parent population.  The average and median composite
spectra should indicate similar trends. Figure~\ref{Fig9} shows this
comparison, which reveals a remarkable similarity between the
composites.  Likewise, there is no apparent difference in \ion{N}{5}
and \ion{O}{1}; both lines are evidently stronger in the high redshift
sources, regardless of the selection technique.  We can conclude that
the color selection methods (primarily BR/BRI) find QSOs with
analogous properties to those of sources found by the SSG grism
survey. Therefore, the corresponding biases, if apparent, affect
the samples to a similar degree.

More useful comparisons may be those with the SDSS sample, which is
claimed to be less biased. According to the authors, the selection
probability in the SDSS detection of high redshift objects does not
strongly depend on the emission line strength, but on the redshift and
the SED shape.  Comparisons between observed EW(Ly$\alpha$)
distributions for our sample, the grism (SSG) sample and the SDSS
survey show that they are consistent with a single distribution
function (probability of being drawn from the same distribution is
respectively KS$_{SDSS/our sample} = 0.633$, KS$_{SDSS/grism} =
0.910$).  The rest and observed frame EW histograms, for both
Ly$\alpha$ and \ion{C}{4}, are displayed in Figure~\ref{Fig8},
separately for color-selected QSOs from our sample, for the
grism-selected objects from SSG, and for the SDSS quasars.  The KS
statistic in each case is consistent with a common parent population.

One possible complication in this comparison may be that the
characteristic luminosities for these different samples are not the
same. Inspection of Figures~\ref{Fig5} and~\ref{Fig6} indicate that
quasars in our sample are somewhat more luminous than those in the
SDSS. The Baldwin Effect would then predict smaller EWs for our
sources.  The fact that the EW distributions are actually similar
could thus be a coincidence, due to a bias favoring strong-lined
objects in our sample. A closer inspection of the luminosity
distributions suggests that such a conspiracy is unlikely to be
significant. There is substantial overlap in luminosities for the
different samples, and differences in their median values are small;
median $L_\nu$ for our QSOs exceeds that of the SDSS sources by only a
factor of 2. 

We conclude that the selection effects, if present, are not more
evident in our sample than in the other high redshift surveys.

\section{Discussion}

Interpretations of the strong cosmological evolution of AGNs are
generally based on the functional form of a (simple) observed
luminosity function of bright QSOs at high $z$ and prescriptions for
the growth and luminosity of the underlying black holes. In the cold
dark matter scenario, as a paradigm for a hierarchical cosmogony that
models the evolution of the luminosity function, quasars are believed
to be a short-lived, first phase of the formation of a galaxy in the
potential well of a dark matter halo \citep{hae93,hai01}.

Observations at high redshifts intercept epochs when quasars are very
young objects, which thus might be expected to have characteristically
low metallicity. However, the trends reported here in the \ion{N}{5}
and \ion{O}{1} broad emission lines (section~\ref{O1N5}) suggest that
the metallicities are typically higher in higher redshift, and
implicitly more luminous QSOs.  The prominent emission lines indicate
that, in spite of their youth, the high-redshift QSOs have undergone
substantial chemical enrichment on a timescale that is short compared
with their life span, implying a rapid star formation process in the
host galaxy.  Therefore, the very existence of objects at $z > 4$
presents severe timing problems (time to form and turn on), that can
be solved only under the assumption that they reside in the most
massive objects which have collapsed at these early epochs, associated
with the rare high peaks in the primordial density field
\citep{tur91,kat94}. This result may naturally arise from the analogy
with the mass-metallicity relationship present in the low $z$ spirals
and ellipticals: massive galaxies reach higher metallicities because
their deeper potential is better able to retain their gas against the
galactic winds built by the thermal pressure from supernovae, while
low-mass systems eject their gas before high enrichments are
attained. Thus, quasar metallicities should be similarly tied to the
gravitational binding energy of the local star-forming regions, and
perhaps also to the total mass of their host galaxies \citep{ham99}.

Consistent with this picture, recent AGN studies have revealed
correlations between the mass of the central black hole and QSO
luminosities, QSO host masses, and the stellar velocity dispersion
\citep{lao98,mag98,fer00,geb00}. Quasar broad lines trace matter
within only the central few parsecs, but the emerging black hole/bulge
connections strongly suggest that the emission-line plasma is closely
linked to the larger star-forming environment.  Recent
cosmic-structure simulations \citep{gne97} show that protogalactic
condensations can form stars and reach higher than solar metallicities
at $z \ga 6$, suggesting that at epochs preceding, or concurrent with
the QSO formation period, the bulge star formation already
occurred. Therefore, interpretation of the abundance data in these
extreme redshift sources yields unique constraints on the evolution of
those environments, indirectly probing the epoch and extent of early
star formation associated with QSOs.  The observed enhancements in the
\ion{N}{5} and \ion{O}{1} emission reinforce the previous evidence of
solar or greater metallicities ($Z > Z_{\sun}$) that occur before
QSOs become observable (Dietrich \& Wilhelm-Erkens 2000; Hamann \&
Ferland 1999 and references therein).

The evidence of structural differences reported in Section ~\ref{O1N5}
for the broad line region in these young sources provides further
motivation for examining their evolutionary context.  If the enhanced
ILR emission is real and not a consequence of selection effects, this
behavior may be an important tracer of accretion physics and its
relationship to redshift and luminosity. Based on our initial reports
of narrow UV lines in $z\ga 4$ QSOs, \citet{mat00} suggested an
analogy between these sources and narrow-line Seyfert 1 (NLS1)
galaxies. In this picture, both classes of objects are in an early
evolutionary phase, in which accretion proceeds at or near the
Eddington limit.  While this scenario has several appealing aspects
for explaining NLS1 phenomena, assigning it in general to high-$z$
QSOs may be premature.  While the latter sources are young in
cosmological terms (ages $\la 2$ Gyr), this does not ensure that they
are in an early phase of accretion, since the timescale for QSO
activity in an individual source is estimated to be only a few times
$10^7$ yr (e.g., Kauffmann \& Haehnelt 2000). Furthermore, narrow
profiles in the UV lines are probably not an adequate basis for
linking our sources to a NLS1 classification, which is based on
optical lines.  As discussed by \citet{wil00}, the UV line widths do
not correlate in a simple way with H${\beta}$ width or with other NLS1
properties.  Further inquiry is needed to understand the implications
of line profile behavior in high-redshift QSOs.

\section{Conclusions}

Although an increasing amount of information is becoming available on
the properties of QSOs discovered at very high redshift, only limited
efforts have been made to survey systematically the emission-line
properties of these objects and/or the selection effects related with
the techniques by which they were discovered.  We have conducted a thorough
analysis of 44 high signal-to-noise spectra from the MMT and Keck
observatories which yield measurements of the most prominent emission
lines in the rest-frame interval 1100{\AA} -- 1700{\AA}.

Composite spectra for the whole data set and for subsets were
constructed and analysed in order to investigate: a) the emission
properties of $z \ga 4$ QSOs in comparison with those of their lower
redshift counterparts, b) the luminosity dependence of the emission
features, and c) the role of selection effects in existing
samples. There are several conclusions that may be inferred from the
work we have outlined here:

1. In terms of their ultraviolet rest-frame spectra, $z \ga 4$ QSOs
strongly resemble quasars at low redshift. Subtle differences are
present, however, and in particular, Ly$\alpha$, \ion{N}{5},
\ion{C}{4} and \ion{O}{1} are stronger in the high $z$ sources.  The
\ion{C}{4} line is also systematically narrower in the $z \ga 4$
objects.

2. Among the high redshift QSOs, a weak Baldwin Effect is possibly
present in Ly$\alpha$ but not in \ion{C}{4} or other lines. 
The lack of a strong trend may reflect the limited span of luminosity
in the existing $z \ga 4$ samples.

3. Selection effects favoring strong-lined objects are a significant
concern for surveys of high redshift QSOs.  Our sample includes several
sources with extremely strong, narrow and peaked lines that may arise
from such a bias.  However, quantitative comparisons with other
existing surveys of $z \ga 4$ QSOs, including those discovered by SDSS,
suggest that our sample overall is not strongly biased in this way.

4. All $z > 4$ composite spectra show strong \ion{N}{5} and
\ion{O}{1}.  The unusual strengths of these features and the narrow
\ion{C}{4} profile suggest the presence of characteristic structural
differences in the BLR, and also high abundances of heavy elements
(solar or up to several times solar metallicities) in quasar
environments at these early times.  These findings are consistent with
standard chemical evolution scenarios for young, massive
bulge-dominated galaxies.

\acknowledgements

We thank Matthias Dietrich, George Djorgovski, Julia Kennefick, and
Lisa Storrie-Lombardi for helpful conversations and for making their
results available in advance of publication.  We are grateful to Don
Schneider for providing us with the SSG spectra in digital form, and
for answering questions concerning grism selection methods.  We thank
our referee, Paul Francis, for valuable comments that resulted in
improvements in the manuscript.  Financial support for this research was
provided by the National Science Foundation through grants AST
98-03072 to C.B.F. and AST 99-84040 to F.H.  The authors wish to
extend special thanks to those of Hawaiian ancestry on whose sacred
mountain we are privileged to be guests.  Without their generous
hospitality, many of the observations presented herein would not have
been possible.

\clearpage

\onecolumn

\clearpage
\begin{figure}
\figurenum{1 continued}
\plotone{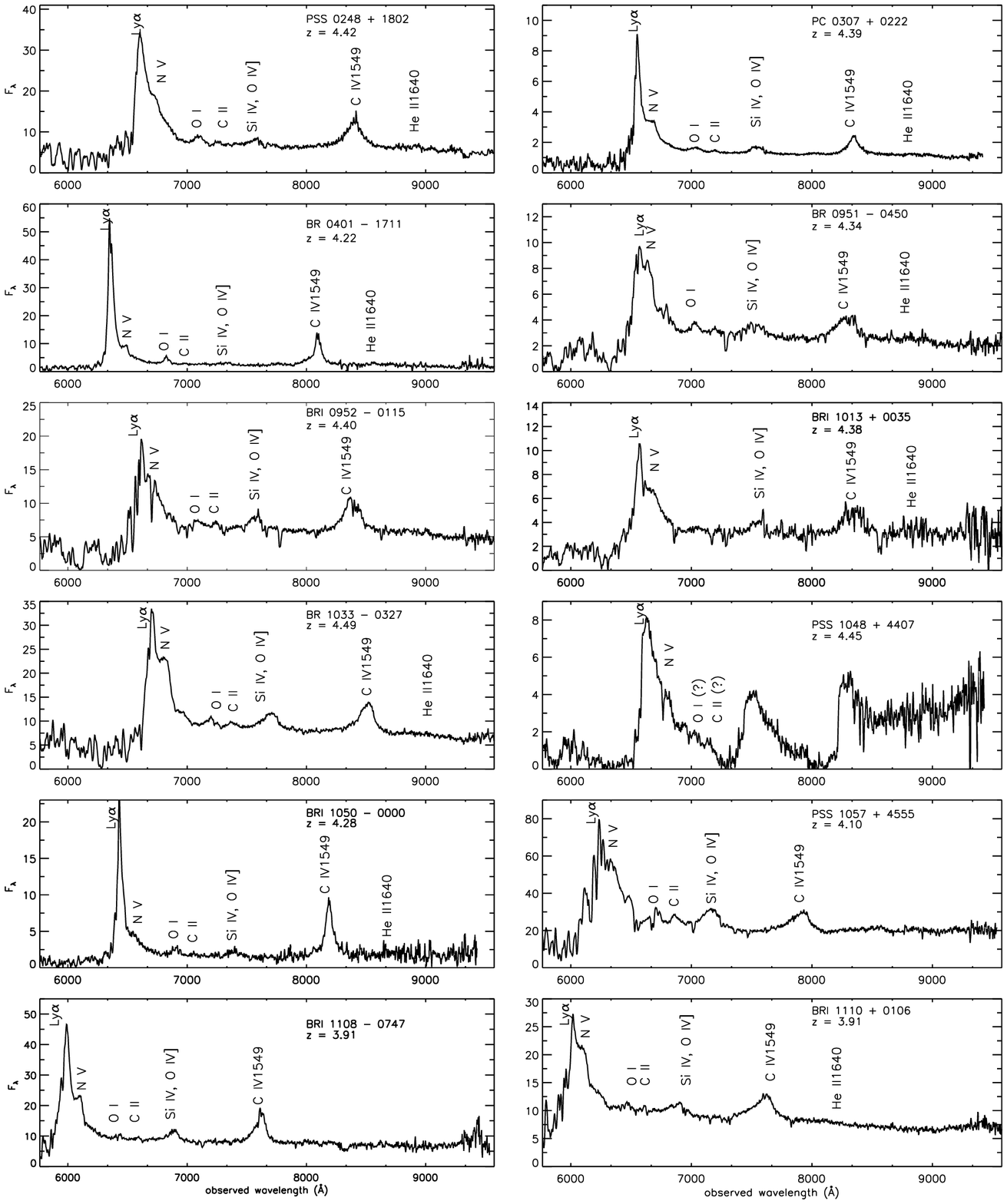}
\caption{ \label{Fig1b}}
\end{figure}

\clearpage
\begin{figure}
\figurenum{1 continued}
\plotone{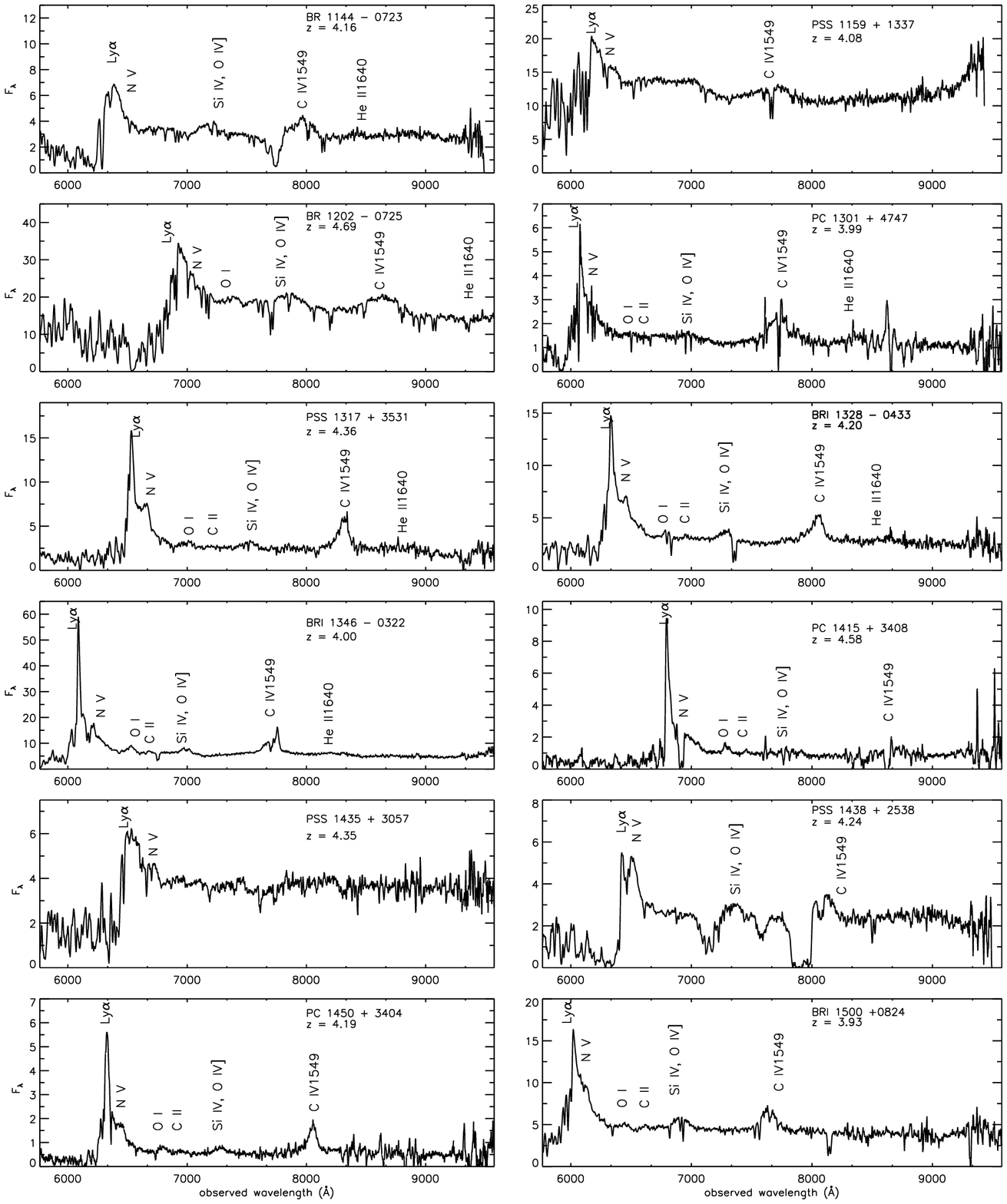}
\caption{ \label{Fig1c}}
\end{figure}

\clearpage
\begin{figure}
\figurenum{1 continued}
\plotone{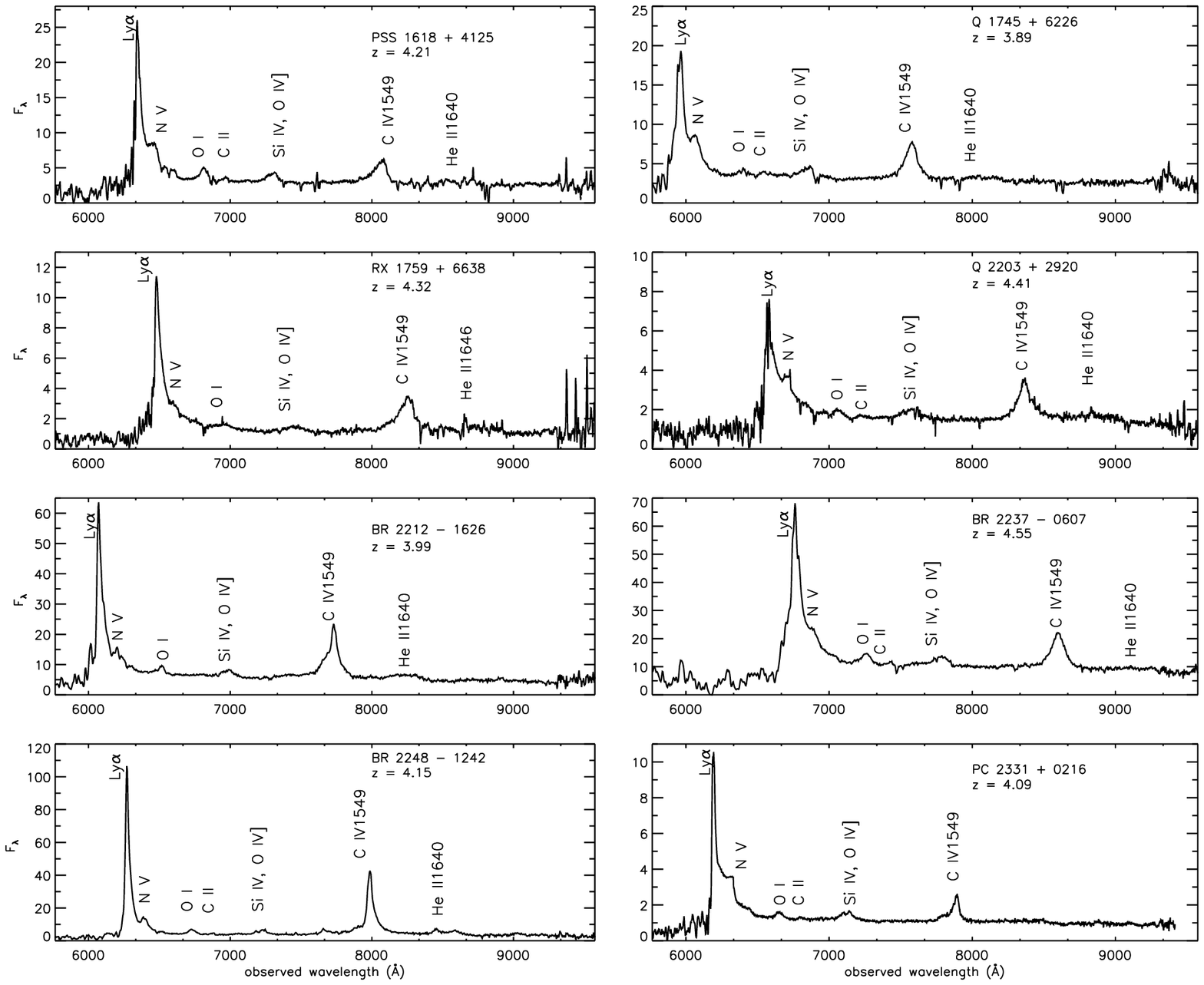}
\caption{ \label{Fig1d}}
\end{figure}

\clearpage

\onecolumn

\begin{figure}
\figurenum{2}
\plotone{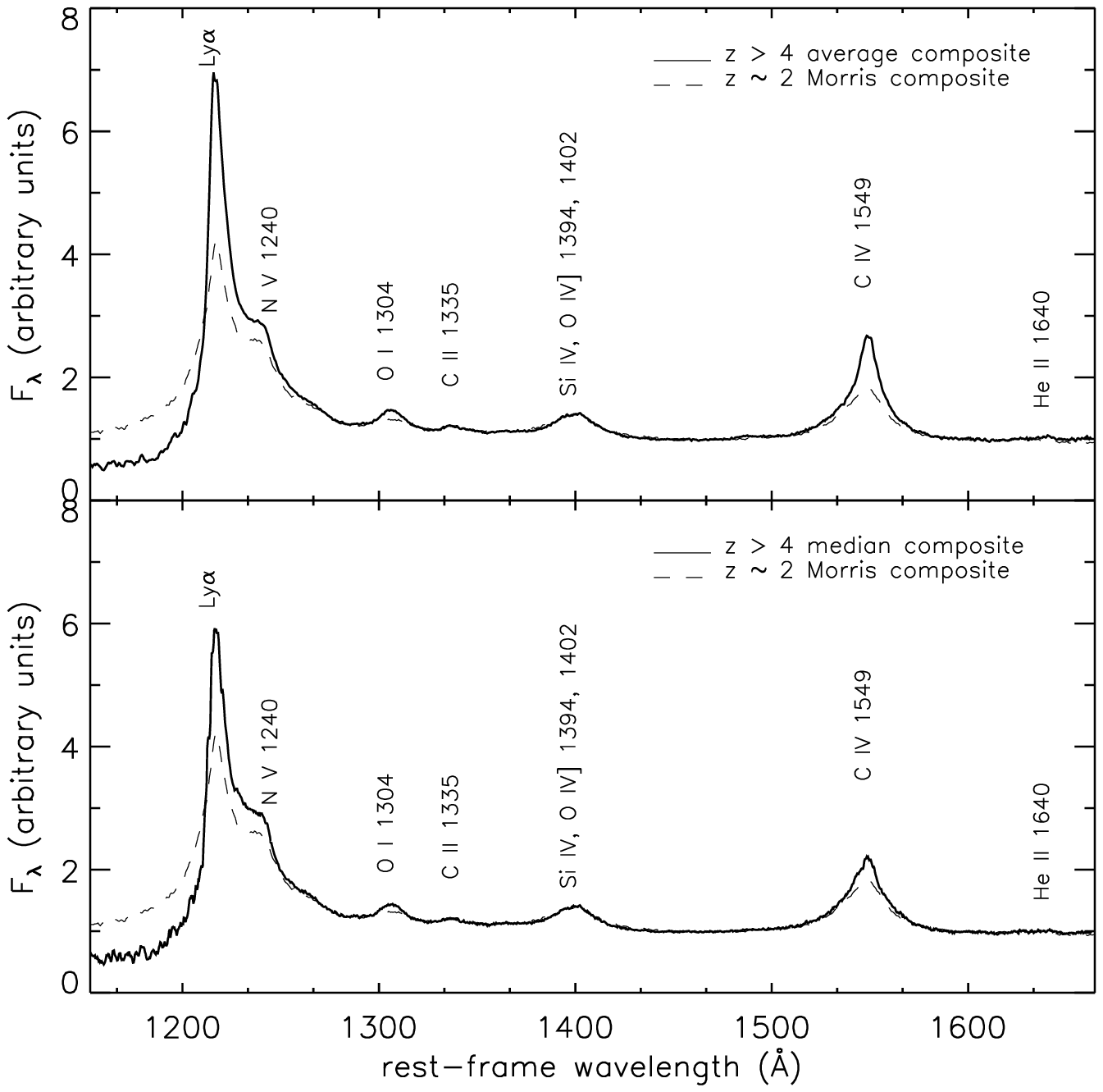}
\caption{Composite spectra (average -upper panel, and median -lower
panel) of $z \ga 4$ QSOs included in this observational program are
shown in comparison with the LBQS composite, representative
of $z \approx 2$ objects. The y-axis has been normalized to unit mean
flux over the wavelength range 1430 \AA -- 1470 \AA. The redshift
measured from the \ion{C}{4} feature was used for the Doppler
correction of the $z \ga 4$ spectra. As a consequence of the different
velocity shifts presented by the low- and high-ionization lines, the
alignment of the high-$z$ spectral features with those in the low $z$
composite required a slight additional shift of our composites.
\label{Fig2}}
\end{figure}

\clearpage
\begin{figure}
\figurenum{3}
\plotone{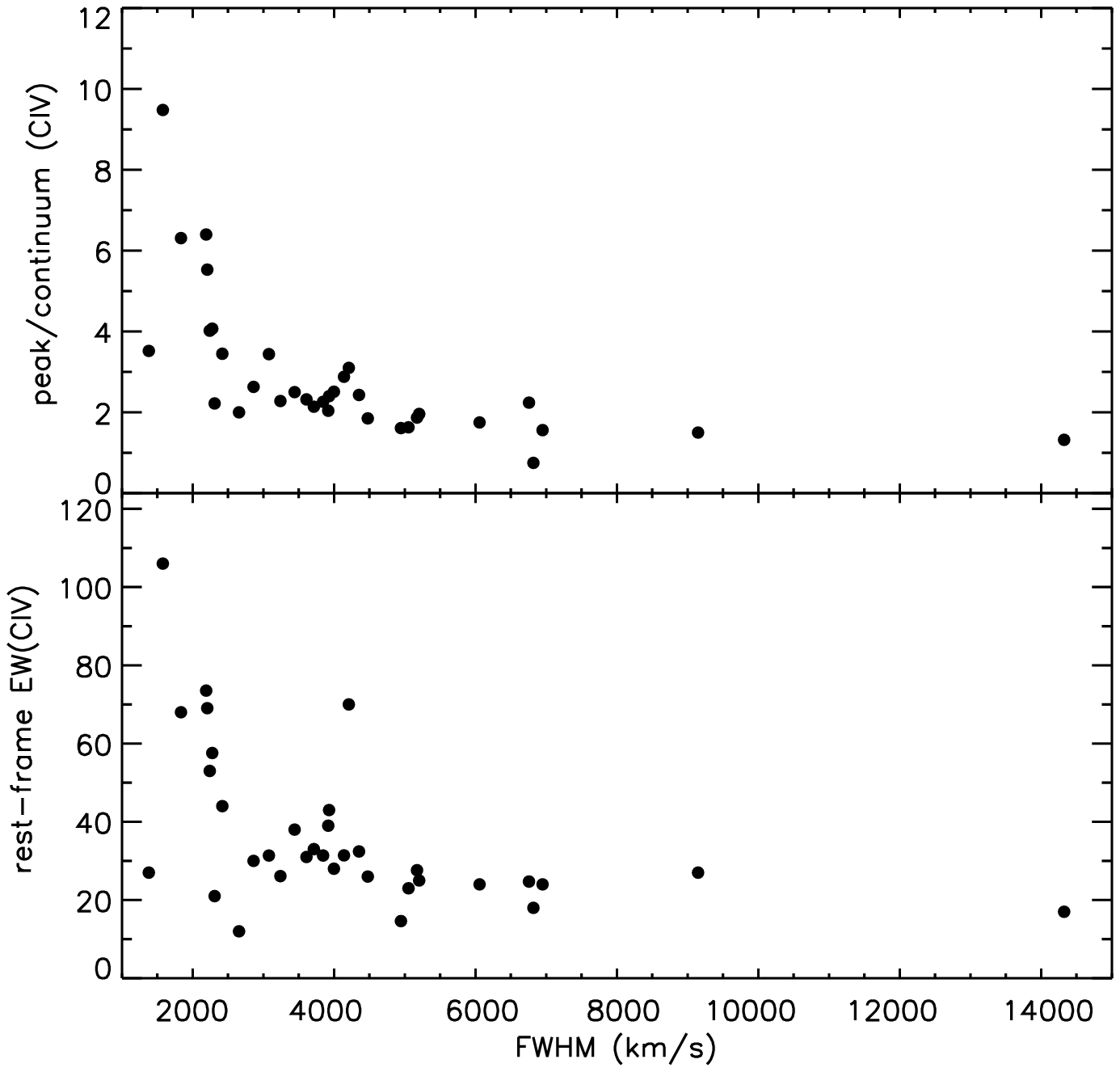}
\caption{Equivalent widths and line peak/continuum ratios for
\ion{C}{4} emission as a function of \ion{C}{4} line widths (full
width at half maximum, FWHM), measured in our sample.  \label{Fig3}}
\end{figure}

\clearpage

\begin{figure}
\figurenum{4}
\plotone{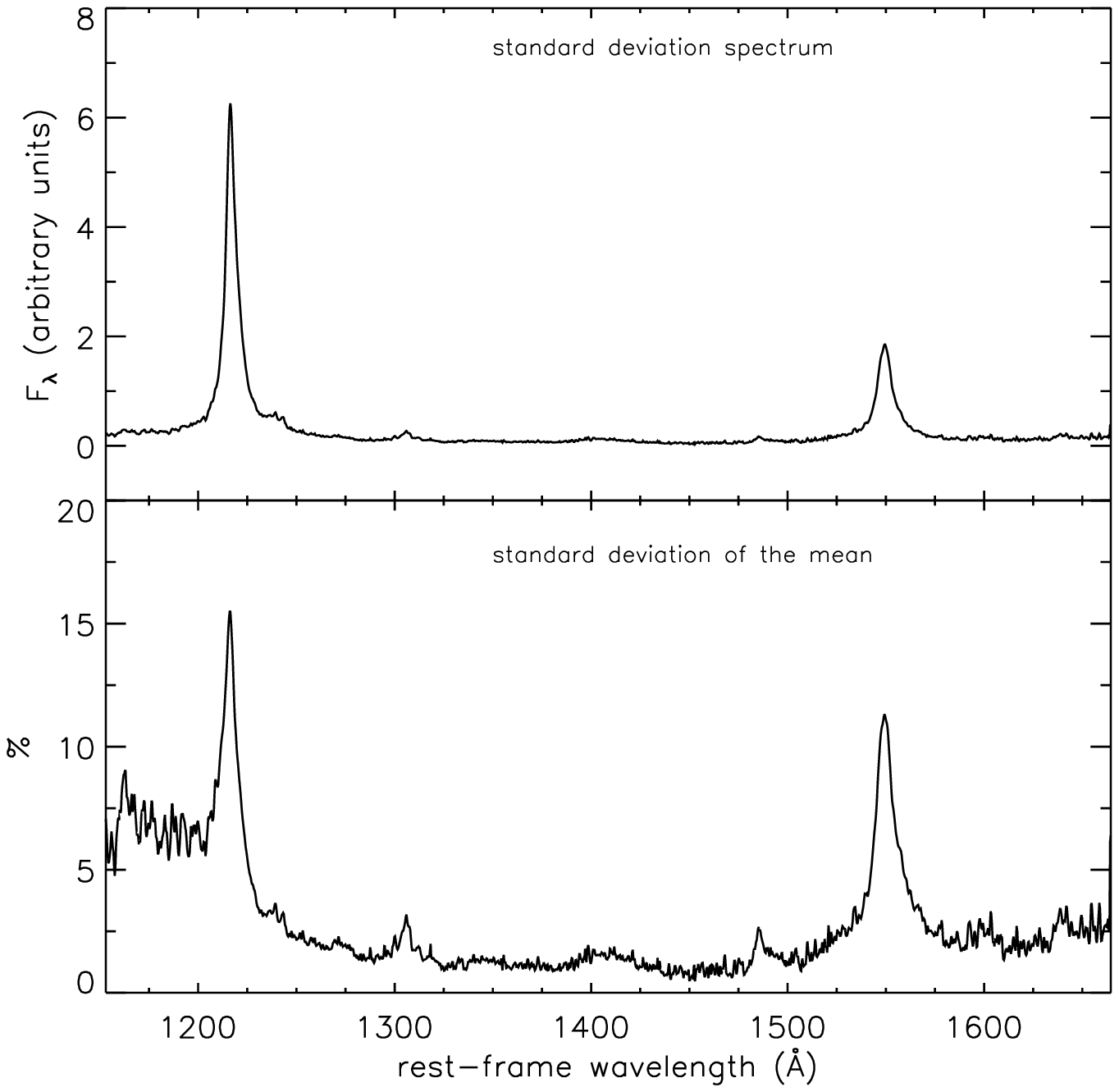}
\caption{Upper panel: Flux standard deviation in F$_{\lambda}$ relative to 
the average composite spectrum, as a function of wavelength for the individual 
spectra comprising the composites. Lower panel: The standard deviation of the 
mean, expressed in \%. \label{Fig4}}
\end{figure}

\clearpage

\begin{figure}
\figurenum{5}
\plotone{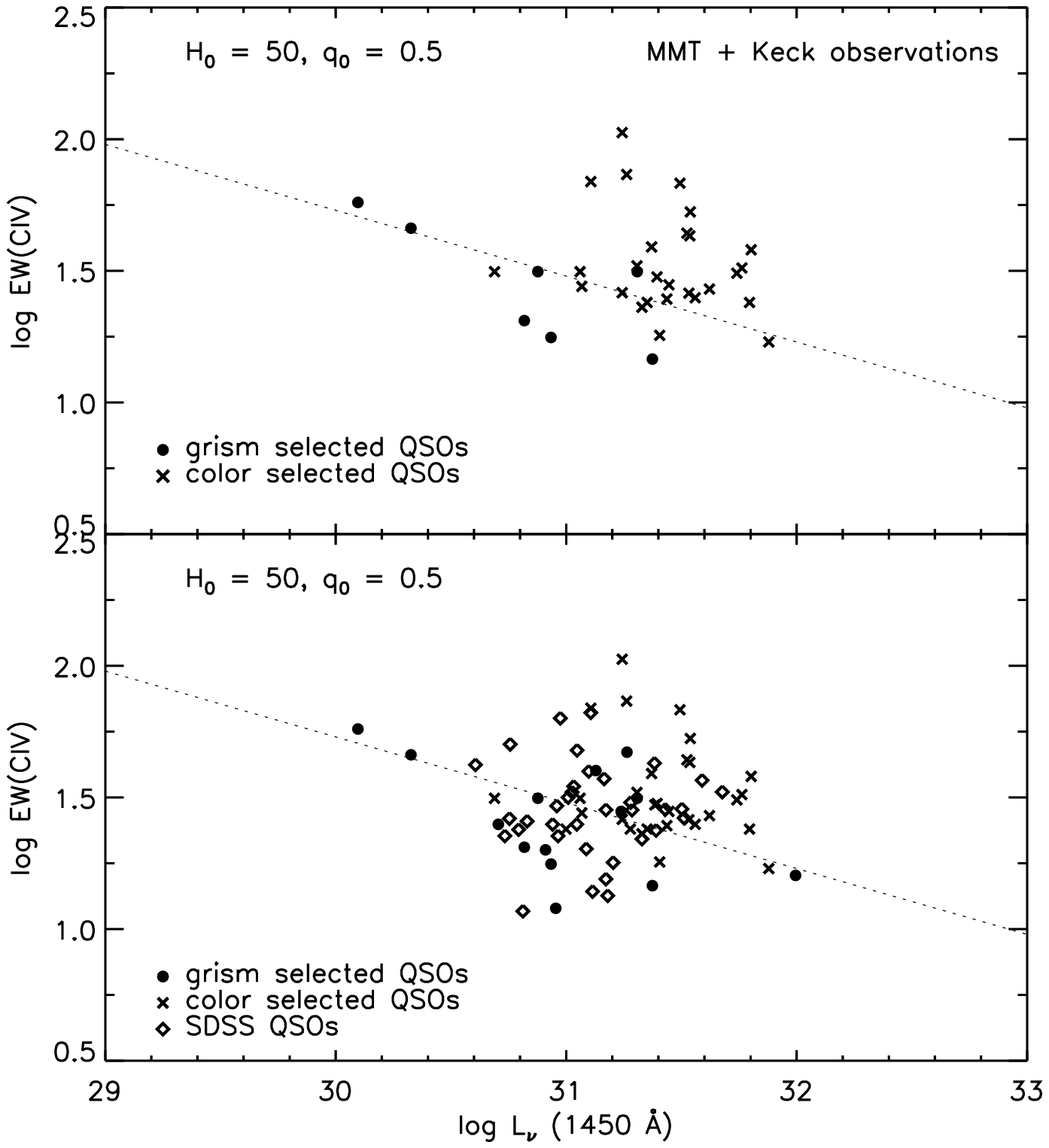}
\caption{Rest-frame EW of the \ion{C}{4} emission line as a function
of absolute luminosity of the quasar (at 1450 \AA\ rest frame). The
Baldwin Effect reported by \citet{osm94}, based on measurements at
lower redshift, is shown by the dotted line. The upper panel shows
measurements from our sample only, with different symbols for the
color ($\times$) and grism ($\bullet$) selected objects. The lower
panel adds measurements for $z \ga 4$ QSOs from \citet{sch91} and the
SDSS survey.  Although consistent with the low redshift trend, the
measurements do not, by themselves, show the expected
correlation. \label{Fig5}}
\end{figure}

\begin{figure}
\figurenum{6}
\plotone{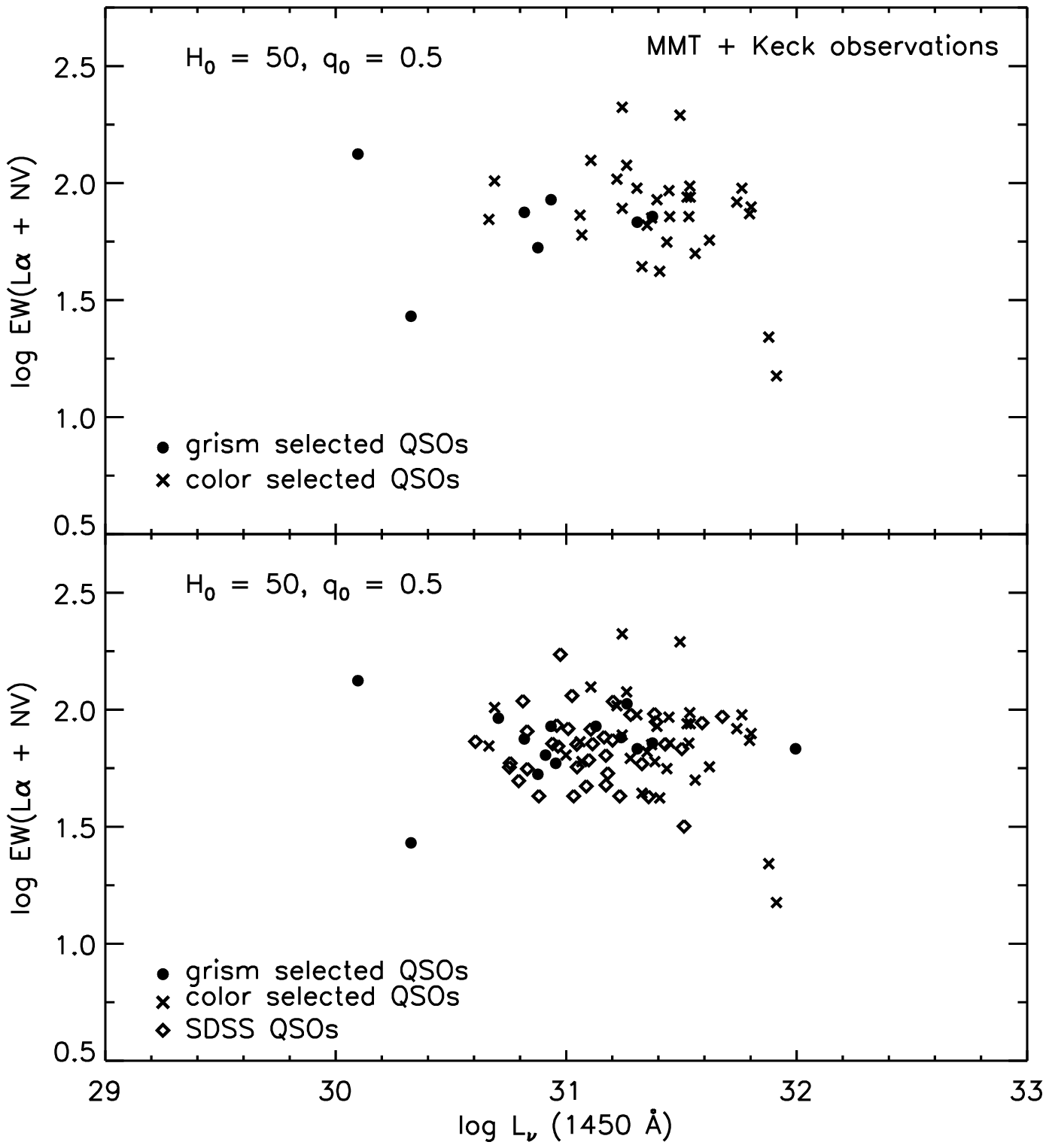}
\caption{Equivalent width of Ly$\alpha$ + \ion{N}{5} versus continuum
luminosity for $z \ga 4$ QSOs. Symbols are the same as in Fig.~\ref{Fig5}. 
\label{Fig6}}
\end{figure}

\clearpage

\begin{figure}
\figurenum{7}
\plotone{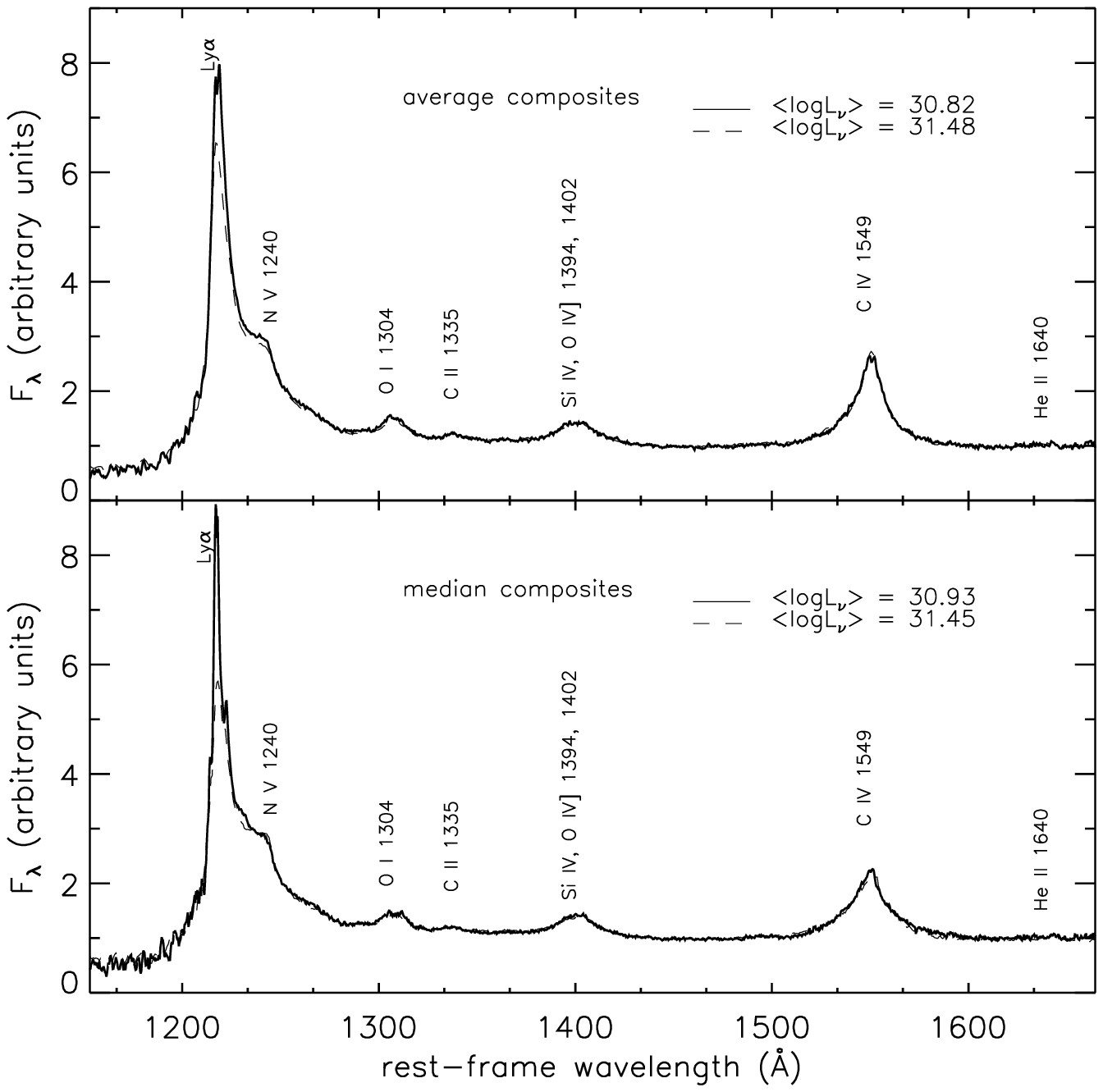}
\caption{Comparison between median and average composites computed for
two sets of quasars, grouped according to their luminosity at 1450\AA\ 
(L$_{\nu}$ expressed in ergs s$^{-1}$ Hz$^{-1}$). The same
normalization as in Figure ~\ref{Fig2} is used. A weak drop in EW with
luminosity can be noted for Ly$\alpha$ but not for \ion{C}{4} or other
lines. \label{Fig7}}
\end{figure}

\clearpage

\begin{figure}
\figurenum{8}
\plottwo{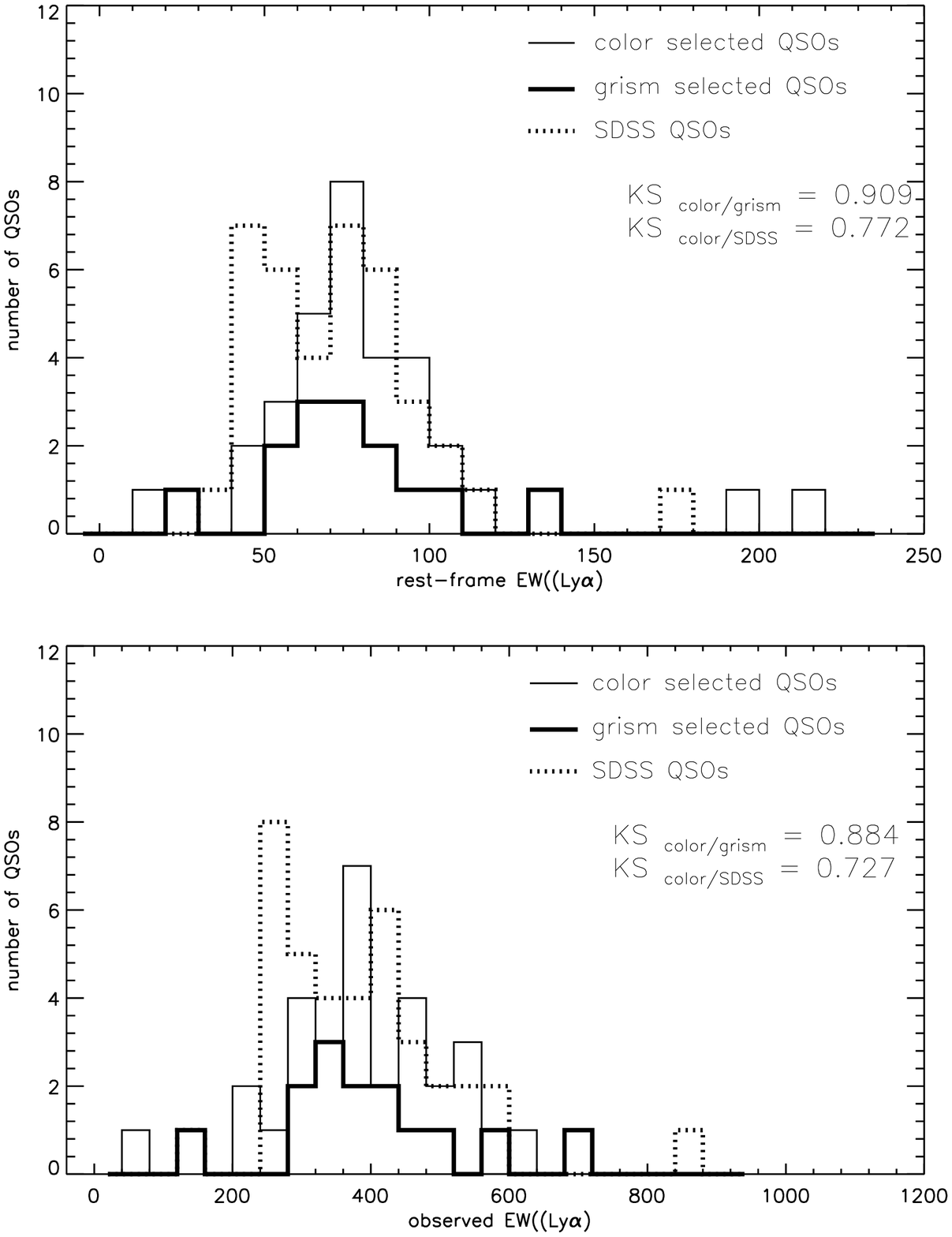}{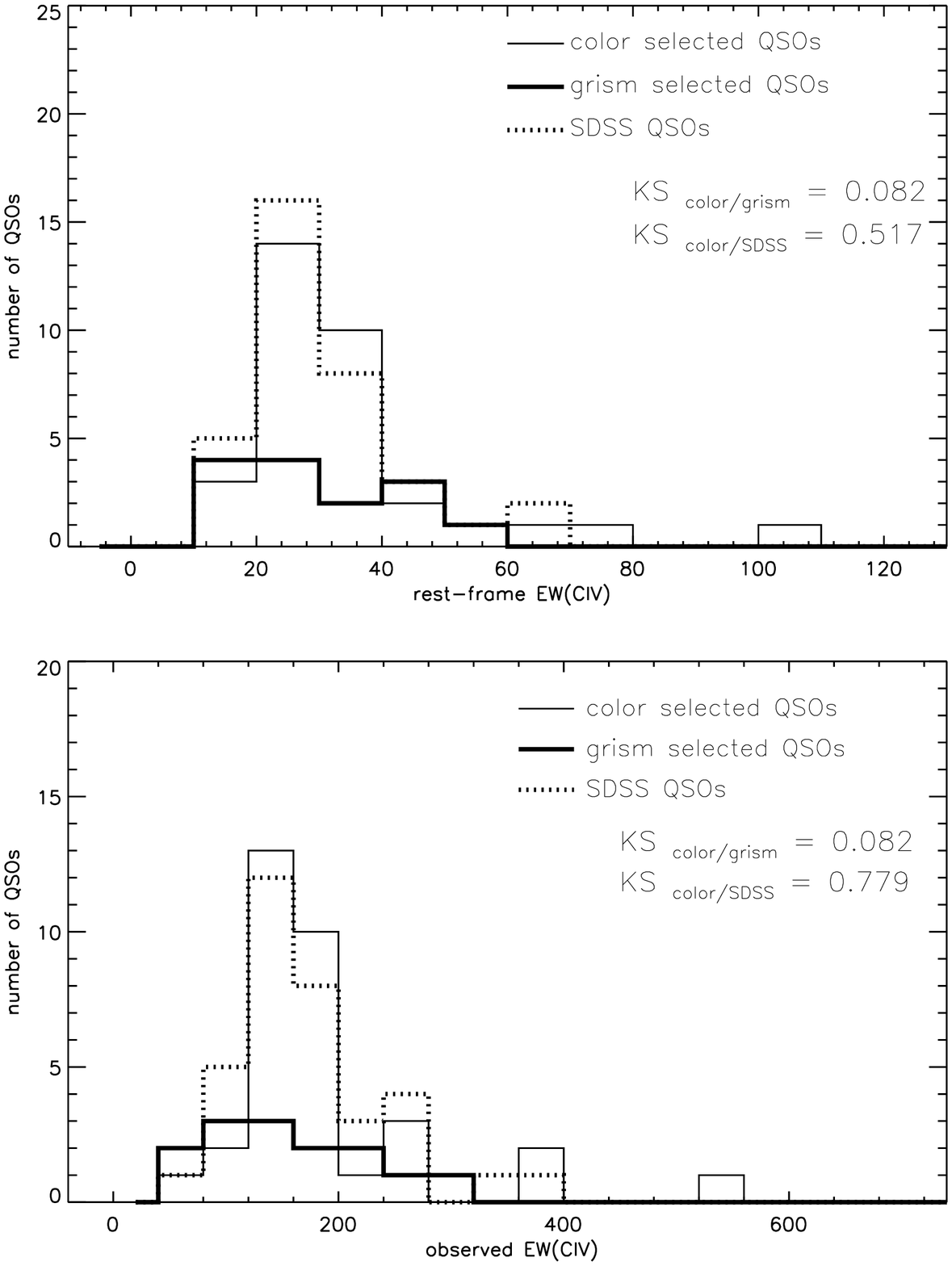}
\caption{ The distributions of the rest-frame (upper panels) and
observed (lower panels) EWs for Ly$\alpha$\ and \ion{C}{4}; three different 
samples are compared using a KS test criterion which indicates that the 
distributions are consistent with a common parent population. \label{Fig8}}
\end{figure}
\clearpage

\begin{figure}
\figurenum{9}
\plotone{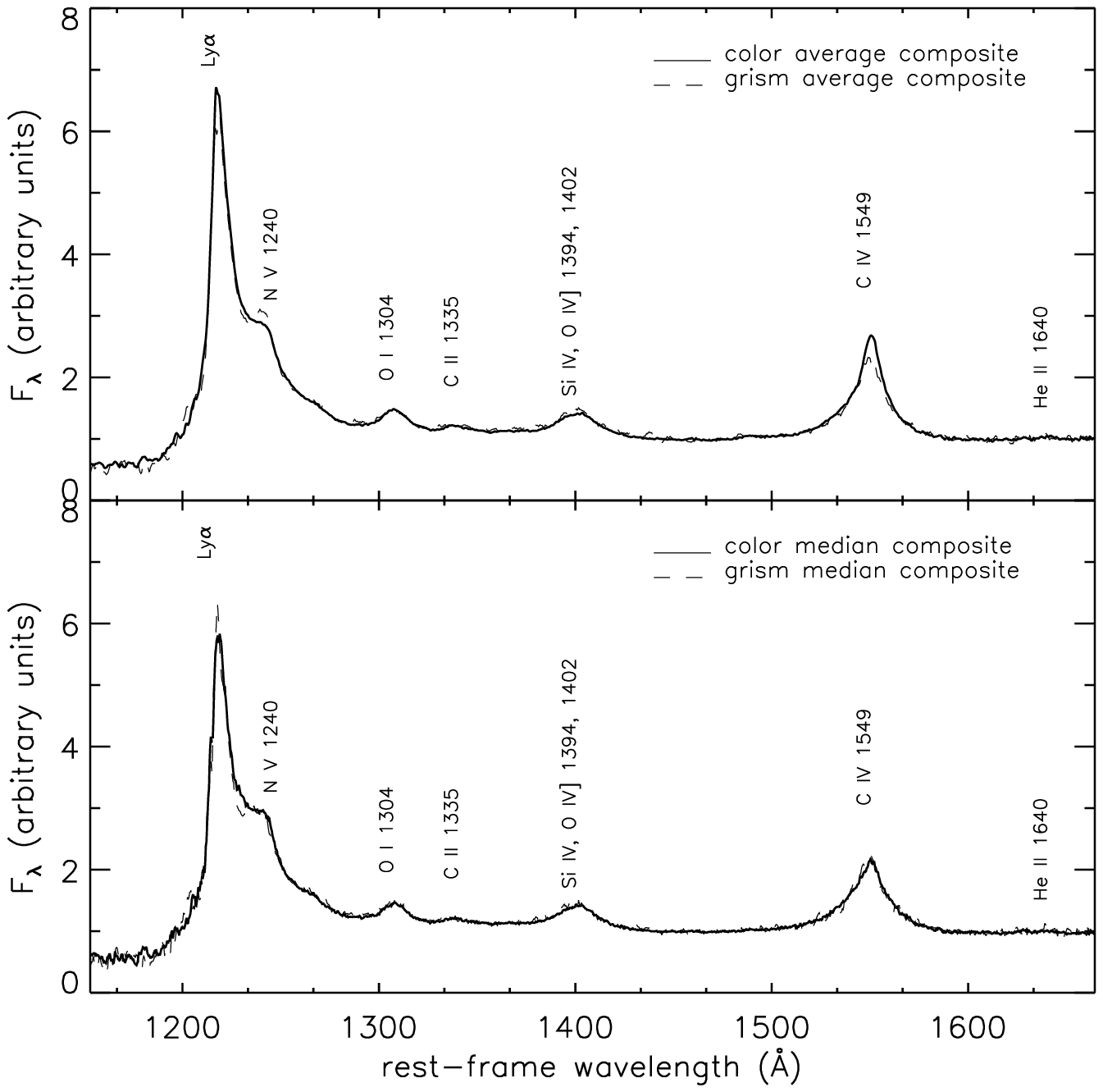}
\caption{Comparison of composite spectra for color-selected and
grism-selected QSOs.  The composites include spectra published by SSG
in addition to the data presented here.  Average and median spectra are
shown in the upper and lower panels, respectively.
\label{Fig9}}
\end{figure}

\clearpage

\begin{deluxetable}{lrrrrcrrrr}
\tablewidth{0pt}
\tablecaption{QSOs observed at MMT and Keck telescopes. \label{tbl-1}}
\tablehead{
\colhead{Object} & \colhead{z\tablenotemark{a}} & \colhead{r 
\tablenotemark{b}} & 
\multicolumn{2}{c}{\ion{C}{4}} & \colhead{}  &  
\multicolumn{1}{c}{Ly$\alpha$ + \ion{N}{5}} & \colhead{log$L_{\nu}$(1450)} &
\colhead{Ref} \\
\cline{4-5} \cline{7-7}\\ 
\colhead{} &  \colhead{} & \colhead{} & \colhead{EW$_1$\tablenotemark{c} 
(\AA)} & \colhead{EW$_2$\tablenotemark{d} (\AA)} & 
\colhead{} &\colhead{EW\tablenotemark{c} (\AA)} & \colhead{} & \colhead{}}
\startdata
PSS 0003+2730 &4.23 &19.3 & 33 & 32 & & 95 & 31.30\tablenotemark{f} & 3\\
BR  0019-1522 &4.52 &19.0 & 43 & 34 & & 97 & 31.53 & 6\\
PC  0027+0521 &4.21 &21.9 & 57 & 43 & &133 & 30.09 & 13 \\
PC  0027+0525\tablenotemark{e} &4.09 &21.5 & -- & -- & & -- & & 13 \\
SDSS 003525+00 &4.75 &21.2 & 69 & 64 & &125 & 31.11 & 4 \\
PSS 0059+0003 &4.15 &19.5 & 26 & 23 & & 78 & 31.24\tablenotemark{f} & 7 \\
BRI 0103+0032 &4.43 &18.6 & 44 & 36 & & 87 & 31.52 & 15\\
PC  0131+0130 &3.78 &19.4 & 17 & 15 & & 72 & 31.37 & 10\\
PSS 0137+2837\tablenotemark{e} &4.30 &19.0 & -- & -- & & -- & & 3 \\
BRI 0151-0025 &4.19 &18.9 & -- & -- & & 72 & 31.44 & 15\\
PSS 0152+0735 &4.05 &19.6 & 27 & 27 & & 60 & 31.06\tablenotemark{f} & 3 \\
BRI 0241-0146 &4.01 &18.2 & 27 & 24 & & 57 & 31.62 & 15 \\
PSS 0248+1802 &4.42 &18.4 & 32 & 31 & & 95 & 31.76\tablenotemark{f} & 8 \\
PC  0307+0222 &4.39 &20.4 & 20 & 20 & & 75 & 30.81 & 11 \\
BR  0401-1711 &4.22 &18.7 & 68 & 61 & &195 & 31.49\tablenotemark{f} & 16 \\
BR  0951-0450 &4.34 &18.9 & 28 & 28 & & 66 & 31.35 & 9, 16 \\
BRI 0952-0115 &4.40 &18.7 & 28 & 27 & & 50 & 31.55 & 16 \\
BRI 1013+0035 &4.38 &18.8 & 36 & 35 & & 56 & 31.43 & 14 \\
BR  1033-0327 &4.49 &18.5 & 26 & 24 & & 72 & 31.53 & 6, 16 \\
PSS 1048+4407\tablenotemark{e} &4.45 &19.3 & -- & -- & & -- & & 8 \\
BRI 1050-0000 &4.28 &18.7 & 73 & 73 & &119 & 31.26 & 14 \\
PSS 1057+4555 &4.10 &17.7 & 24 & 21 & & 74 & 31.79 & 8 \\
BRI 1108-0747 &3.91 &18.1 & 28 & 25 & & 93 & 31.44 & 16 \\
BRI 1110+0106 &3.91 &18.3 & 20 & 19 & & 42 & 31.40 & 16 \\
BR  1144-0723\tablenotemark{e} &4.16 &18.6 & -- & -- & & -- & & 16 \\
PSS 1159+1337 &4.08 &18.5 & -- & -- & & 15 & 31.91\tablenotemark{f} & 3 \\
BR  1202-0725 &4.69 &18.7 & 17 & 17 & & 22 & 31.87 & 6, 16 \\
PC  1301+4747 &3.99 &21.3 & 46 & 46 & & 27 & 30.32 & 12 \\
PSS 1317+3531 &4.36 &19.1 & 30 & 29 & & 85 & 31.39\tablenotemark{f} & 8 \\
BRI 1328-0433 &4.20 &19.0 & 39 & 26 & & 71 & 31.37 & 9, 16 \\
BRI 1346-0322 &4.00 &18.8 & -- & -- & &104 & 31.21 & 16 \\
PC  1415+3408 &4.58 &21.4 & -- & -- & & 70 & 30.66 & 13 \\
PSS 1435+3057 &4.35 &19.3 & -- & -- & & 20 &  & 7 \\
PSS 1438+2538\tablenotemark{e} &4.24 &19.5 & -- & -- & & -- & & 8 \\
PC  1450+3404 &4.19 &20.8 & 44 & 38 & & 102& 30.68 & 13 \\
BRI 1500+0824 &3.93 &18.8 & 23 & 18 & & 44 & 31.32 & 15 \\
PSS 1618+4125 &4.21 &19.6 & 31 & 26 & & 73 & 31.06\tablenotemark{f} & 3 \\
Q   1745+6226 &3.89 &18.8 & 38 & 34 & & 79 & 31.80\tablenotemark{f} & 1 \\
RX  1759+6638 &4.32 &20.0 & 45 & 40 & & 68 & 31.30 & 5 \\
Q   2203+2900 &4.41 &20.8 & 31 & 26 & & 53 & 30.87 & 2 \\
BR  2212-1626 &3.99 &18.1 & 51 & 50 & & 87 & 31.53 & 16 \\
BR  2237-0607 &4.55 &18.3 & 31 & 26 & & 83 & 31.74\tablenotemark{f} & 6, 16 \\
BR  2248-1242 &4.15 &18.5 &106 & 98 & &211 & 31.24 & 6, 16 \\
PC  2331+0216 &4.09 &20.0 & 17 & 15 & & 85 & 30.93 & 11  \\
\enddata
\tablenotetext{a}{calculated from our spectra using the 
\ion{C}{4} emission-line} 
\tablenotetext{b}{photometry in $r$
or analogous bandpass (R for the APM survey, $r$ for PSS, $r^{\ast}$ for
SDSS, $r_4$ for SSG) as reported in the respective discovery papers} 
\tablenotetext{c}{rest-frame EWs calculated by direct integration of the 
line flux} 
\tablenotetext{d}{rest-frame EWs calculated by fitting the line to a 
single Gaussian}
\tablenotetext{e}{exhibits broad absorption lines (BAL); $z$ values are taken 
from the discovery papers} 
\tablenotetext{f}{values derived using the flux $f_{\lambda}(1450)$ 
measured from our spectra; estimated errors are approximately $\pm$ 0.2 dex.}

\tablerefs{
(1) \citet{beck92}; (2) Dickinson \& McCarthy 1987;
(3) S. Djorgovski, private comunication; (4) Fan et al. 1999; 
(5) Henry et al. 1994; (6) Isaak et al. 1994;  (7) Kennefick et al. 1995a; 
(8) Kennefick et al. 1995b; (9) Kennefick et al. 1996;
(10) Schmidt et al. 1987; (11) Schneider et al. 1989; 
(12) Schneider et al. 1991; (13) Schneider et al. 1997; 
(14) Smith et al. 1994a; (15) Smith et al. 1994b; 
(16) Storrie-Lombardi et al. 1996.}
\end{deluxetable}

\clearpage


\begin{thebibliography}{}
\bibitem[Baldwin (1977)]{bal77} Baldwin, J.A. 1977, \apj, 214, 769 
\bibitem[Baldwin et al. (1996)]{bal96} Baldwin, J.A., et al. 1996, \apj, 461, 
664
\bibitem[Becker et al. (1992)]{beck92} Becker, R. H., Helfand, David J., \&
 White, R. L. 1992, \aj, 104, 531
\bibitem[e. g., Brotherton et al. (1994)]{bro94} Brotherton, M. S., 
Wills, B. J., Francis, P. J., \& Steidel, C. C. 1994, \apj, 430, 495
\bibitem[Brotherton et al. (2001)]{bro01} Brotherton, M. S., 
Tran, Hien D., Becker, R. H., Gregg, Michael D., Laurent-Muehleisen, S. A., 
\& White, R. L. 2001, \apj, 546, 775
\bibitem[Cardelli, Clayton \& Mathis (1989)]{car89} Cardelli, J. A., Clayton, 
G. C., \& Mathis, J. S. 1989, \apj, 345, 245
\bibitem[Dickinson \& McCarthy (1987)]{dick87} Dickinson, M., \& McCarthy, 
P. J. 1987, \baas, 19, 1125
\bibitem[Dietrich \& Wilhelm-Erkens (2000)]{die00} Dietrich, M., \& 
Wilhelm-Erkens, U. 2000, A\&A, 354, 17 
\bibitem[Dietrich et al. (2001)]{die01} Dietrich, M., et al. 2001, in 
preparation 
\bibitem[Dumont \& Mathez (1981)]{dum81} Dumont, A. M., \& Mathez, G. 1981, 
A, \&A 102, 1
\bibitem[Espey et al. (1989)]{esp89} Espey, B. R., Carswell, R. F., 
Bailey, J. A., Smith, M. G., \& Ward, M. J. 1989, \apj, 342, 666
\bibitem[SDSS; Fan et al. (1999)]{fan99} Fan, X., et. al 1999, \aj, 
118, 1
\bibitem[SDSS; Fan et al. (2000)]{fan00} Fan, X., et al. 2000, \aj, 
119, 1
\bibitem[SDSS; Fan et al. (2001)]{fan00b} Fan, X., et al. 2001, \aj, 121, 54 
\bibitem[Ferland et al. (1996)]{fer96} Ferland, G. J., et al. 1996, \apj, 
461, 664
\bibitem[Ferrarese \& Merritt (2000)]{fer00} Ferrarese, L., \& Merritt, D. 
2000, \apj, 539, L9
\bibitem[Filippenko (1982)]{fil82} Filippenko, A. V. 1982, \pasp, 94, 715
\bibitem[Forster et al. (2000)]{for00} Forster, K., Paul J. G., 
Aldcroft T. L., Vestergaard M., Foltz G. B., \& Hewett P. C. 2001, \apjs, 
in press
\bibitem[Francis et al. (1991)]{fra91} Francis, P. J., Hewett, P. C.,
Foltz, C. B., Chaffee, F. H., Weymann, R. J., \& Morris, S. L. 1991, \apj, 
373, 465
\bibitem[Francis et al. (1992)]{fra92} Francis, P. J., Hewett, P. C.,
Foltz, C. B., \& Chaffee, F. H. 1992, \apj, 398, 476F
\bibitem[Francis, Hooper \& Impey (1993)]{fra93} Francis, P. J., Hooper, E. J.,
 \& Impey, C. D. 1993, \aj, 106, 417
\bibitem[Gaskell (1982)]{gas82} Gaskell, C. M. 1982, \apj, 263, 79
\bibitem[Gebhardt et al. (2000)]{geb00} Gebhardt, K., et al. 2000, \apj, 539, 
L13 
\bibitem[Gnedin \& Ostriker (1997)]{gne97}  Gnedin, N. Y., \& Ostriker, J. P. 
1997, \apj, 486, 581
\bibitem[Haehnelt \& Rees (1993)]{hae93} Haehnelt, M. G., \& Rees, M, J. 1993, 
\mnras, 263, 168
\bibitem[Haiman \& Loeb (2001)]{hai01} Haiman, Z., \& Loeb, A. 2001, in press
\bibitem[Hamann \& Ferland (1992)]{ham92} Hamann, F., \& Ferland, G. 1992, 
\apj, 391, L53
\bibitem[Hamann \& Ferland (1993)]{ham93} Hamann, F., \& Ferland, G. 1993, 
\apj, 418, 11
\bibitem[Hamann \& Ferland (1999)]{ham99} Hamann, F., \& Ferland, G. 1999 
\araa, 37, 487 
\bibitem[Henry et al. (1994)]{hen94} Henry, J. P., et al. 1994, \aj, 107, 127
\bibitem[Irwin et al. (1991)]{irw91} Irwin, M., McMahon, R. G., \&
Hazard, C.  1991, in ASP Conf.  Ser. 21, The Space Distribution of
Quasars, ed. D. Crampton (San Francisco: ASP), 117
\bibitem[Isaak et al. (1994)]{isa94} Isaak K. G., McMahon, R. G., Hills, R. E.,
 \& Withington, S. 1994, \mnras, 269, 28
\bibitem[Katz et al. (1994)]{kat94} Katz, N., Quinn, T., Bertschinger, E., 
\& Gelb, J. M. 1994, \mnras, 270, L71
\bibitem[Kauffmann \& Haehnelt (2000)]{kau00} Kauffmann, G., \& Haehnelt, M. 
2000, \mnras, 311, 576
\bibitem[Kennefick et al. (1995a)]{ken95a} Kennefick, J. D., et al. 1995a, 
\aj, 110, 78
\bibitem[Kennefick et al. (1995b)]{ken95b} Kennefick, J. D., Djorgovski, S. G.,
 \& de Carvalho, R. R., 1995b, \aj, 110, 2553
\bibitem[Kennefick et al. (1996)]{ken96} Kennefick, J. D., Djorgovski, S. G., 
\& Meylan, G., 1996, \aj, 111, 1816
\bibitem[Korista et al. (1998)]{kor98} Korista, K., Baldwin, J., \& Ferland, G. 
1998, \apj, 507, 24 
\bibitem[Laor (1998)]{lao98} Laor, A. 1998, \apj, 505L, 83L
\bibitem[Magorrian et al. (1998)]{mag98} Magorrian, J., et al. 1998, 
\aj, 115, 2285
\bibitem[Mathur (2000)]{mat00} Mathur S. 2000, \mnras, 314, 17 
\bibitem[McIntosh et al. (1999)]{mci99} McIntosh, D. H., Rix, H.-W.,
 Rieke, M. J., \& Foltz, C. B. 1999, \apj, 517, 73
\bibitem[Netzer (1990)]{net90} Netzer, H. 1990, in Active Galactic Nuclei, ed. 
T. J.-L. Courvoisier, \& M. Mayor (Berlin: Springer), 57
\bibitem[Netzer \& Penston (1976)]{net76} Netzer, H., \& Penston, M. V. 1976, 
\mnras, 174, 319
\bibitem[Oke et al. (1995)]{oke95} Oke, J. B., et al. 1995, \pasp, 107, 375
\bibitem[Osmer, Porter \& Green (1994)]{osm94} Osmer, P. S., Porter, A. C., 
\& Green, R.F. 1994, ApJ, 436, 678
\bibitem[Osmer \& Shields (1999)]{osm99} Osmer, P. S., \& Shields,
J. C.  1999, ASP Conf. Ser.162, Quasars and Cosmology, ed. G. Ferland, 
\& J. Baldwin (San Francisco: ASP), 235
\bibitem[Shields \& Hamann (1997)]{shi97} Shields, J. C., \& Hamann, F. 1997, 
RevMexAA (Serie de Conf.), 6, 221
\bibitem[Schlegel, Finkbeiner \& Davis (1998)]{sch98} Schlegel, D. J., 
Finkbeiner, D. P., \& Davis, M. 1998, \apj, 500, 525
\bibitem[Schmidt, Weymann \& Foltz (1989)]{schm89} Schmidt, G. D., Weymann, 
R. J., \& Foltz, C. B. 1989, \pasp, 101, 713 
\bibitem[Schmidt, Schneider \& Gunn (1987)]{sch87} Schmidt, M., Schneider, 
D. P., \& Gunn, J. E. 1987, \apj, 316, L1
\bibitem[Schmidt, Schneider \& Gunn (1995)]{sch95} Schmidt, M., Schneider, 
D. P., \& Gunn, J. E. 1995, \aj, 110, 68S
\bibitem[Schneider, Schmidt \& Gunn (1989)]{sch89} Schneider, D. P., Schmidt, 
M., \& Gunn, J. E. 1989, \aj, 98, 1507
\bibitem[Schneider, Schmidt \& Gunn (1991)]{sch91} Schneider, D. P., Schmidt, 
M., \& Gunn, J. E. 1991, \aj, 101, 2004
\bibitem[Schneider, Schmidt \& Gunn (1997)]{sch97} Schneider, D. P., Schmidt, 
M., \& Gunn, J. E. 1997, \aj, 114, 36
\bibitem[Smith et al. (1994a)]{smi94a} Smith, J. D., Thompson, D., \& 
Djorgovski, S. G. 1994a, \aj, 107, 24
\bibitem[Smith et al. (1994b)]{smi94b} Smith, J. D., et al. 1994b, \aj, 108, 
1147 
\bibitem[Storrie-Lombardi et al. (1996)]{sto96} Storri-Lombardi, L. J., 
McMahon, R. G., Irwin, M. J., \& Hazard, C. 1996, \apj, 468, 121S
\bibitem[Storrie-Lombardi et al. (2001)]{sto01} Storri-Lombardi, L. J., 
Irwin, M. J., McMahon, R. G., \& Hook, I. M. 2001, \mnras, 322, 933S
\bibitem[Turner (1991)]{tur91} Turner, E. L. 1991, \aj, 101, 1
\bibitem[Tytler \& Fan (1992)]{tyt92} Tytler, D., \& Fan, X. M. 1992, 
\apjs, 79, 1
\bibitem[Verner et al. (1996)]{ver96} Verner, D. A., Verner, E. M., \& Ferland, 
G. J. 1996, Atomic Data Nucl. Data Tables, 64, 1
\bibitem[Wills et al. (1993)]{wil93} Wills, B. J., Brotherton, M. S., 
Fang, D., Steidel, C. C., \& Sargent, W. L. W. 1993, \apj, 415, 563
\bibitem[Wills et al. (2000)]{wil00} Wills, B. J., Shang, Z., \& Yuan J. M. 
2000, New Astronomy Reviews, 44, 511
\end{thebibliography}
\end{document}